\documentclass{aa}

\usepackage{graphics}
\usepackage{epsfig}

\def\bib{\bibitem{}}

\newcommand{\xia}{\overline{\xi}}
\newcommand{\xib}{\overline{\xi}}

\newcommand{\gam}{\gamma}
\newcommand{\inta}{\int_{-i\infty}^{+i\infty}}

\newcommand{\be}{\begin{equation}}
\newcommand{\ee}{\end{equation}}

\newcommand{\ba}{\begin{eqnarray}}
\newcommand{\ea}{\end{eqnarray}}

\newcommand{\lag}{\langle}
\newcommand{\rag}{\rangle}
\newcommand{\De}{{\cal D}}
\newcommand{\om}{\omega}

\newcommand{\kappamin}{\kappa_{\rm min}}

\newcommand{\kpar}{k_{\parallel}}
\newcommand{\Gam}{\Gamma}

\newcommand{\Om}{\Omega_{\rm m}}
\newcommand{\Ol}{\Omega_{\Lambda}}

\newcommand{\Map}{M_{\rm ap}}
\newcommand{\Maph}{\hat{M}_{\rm ap}}

\newcommand{\oma}{\overline{\omega}}
\newcommand{\wh}{\hat{w}}

\def\mF{{\cal F}}
\def\vq{{\bf q}}

\def\vr{{\bf r}}
\def\vx{{\bf x}}
\def\vk{{\bf k}}
\def\vl{{\bf l}}
\def\d{{\rm d}}
\def\ii{{\rm i}}
\def\mg{\langle}
\def\md{\rangle}

\def\deltap{\delta_{\rm proj.}}
\def\deltapt{\delta_{{\rm proj.},\theta}}
\def\drad{{\cal R}}

\def\varphip{\varphi_{\rm proj.}}
\def\varphic{\varphi_{\rm cyl.}}
\def\ka{\kappa}

\begin{document}
%
% US additions
% 
%\topmargin=2.5 cm
%\evensidemargin=2.5 cm
%\oddsidemargin=2.5 cm
%\thispagestyle{empty}
%
%
\renewcommand{\textfraction}{.01}
\renewcommand{\topfraction}{0.99}
\renewcommand{\bottomfraction}{0.99}
\setlength{\textfloatsep}{2.5ex}
\thesaurus{Sect.02 (12.03.4; 12.07.1; 12.12.1)}
\title{Construction of the one-point PDF of the local aperture mass in weak
lensing maps }   
\author{Francis Bernardeau \& Patrick Valageas}
\institute{Service de Physique Th\'eorique, 
CE de  Saclay, 91191 Gif-sur-Yvette, France}
\date{Received / Accepted }
\maketitle
\markboth{F. Bernardeau \& P. Valageas}{Construction of the one point PDF of
the local aperture mass}

\begin{abstract}

We  present  a  general  method  for the  reconstruction  of  the  one-point
Probability Distribution Function of the local aperture mass in weak lensing
maps. Exact results, that neglect  the lens-lens coupling and departure form
the  Born approximation,  are derived  for  both the  quasilinear regime  at
leading  order   and  the  strongly  nonlinear  regime   assuming  the  tree
hierarchical model is valid.  We  describe in details the projection effects
on the  properties of  the PDF and  the associated generating  functions. In
particular, we  show how the generic  features which are common  to both the
quasilinear  and  nonlinear  regimes  lead  to  two  exponential  tails  for
$P(\Map)$. We briefly  investigate the dependence of the  PDF with cosmology
and with the shape of the  angular filter. Our predictions are seen to agree
reasonably well with the results of numerical simulations and should be able
to serve as foundations for  alternative methods to measure the cosmological
parameters that take advantage of the full shape of the PDF.

\end{abstract}


\section{Introduction}

Recent reports of  cosmic shear detection (van Waerbeke  et al. 2000a, Bacon,
Refregier  \& Ellis 2000,  Wittman et  al. 2000,  Kaiser, Wilson  \& Luppino
2000) have  underlined the  interest  that such  observations  can have  for
exploring the large-scale structures  of the Universe.  Previous papers have
stressed that not  only it could be possible to  measure the projected power
spectrum (Blandford et al. 1991,  Miralda-Escud\'e 1991, Kaiser 1992) of the
matter  field, but  also that  non-linear effects  could be  significant and
betray  the  value   of  the  density  parameter  of   the  Universe.   More
specifically, Bernardeau,  van Waerbeke and  Mellier (1997) have  shown that
the  skewness, third  order  moment  of the  local  convergence field,  when
properly  expressed in terms  of the  second moment  can be  a probe  of the
density   parameter  independently   of   the  amplitude   of  the   density
fluctuations. This  result can be extended  to higher order  moments, to the
nonlinear  regime (Jain  \& Seljak  1997, Hui  1999, Munshi  \&  Coles 2000,
Munshi \&  Jain 1999b) and  the whole shape  of the one-point  PDF (Valageas
2000a,b; Munshi \& Jain 1999a,b).

In  case  of  weak lensing  surveys  it  appears  however  that it  is  more
convenient  to  consider  the   so  called  aperture  mass  statistics  that
corresponds to  filtered convergence fields with a  compensated filter, that
is with  a filter of  zero spatial average  (Kaiser et al.   1994, Schneider
1996).  Indeed,  it is  possible to  relate the local aperture mass to the
observed shear  field only, whereas,  in contrast, convergence  maps require
the resolution of  a non-local inversion problem and are  only obtained to a
mass sheet degeneracy.  The aperture mass statistics have proved valuable in
particular for cosmic variance related issues (Schneider et al. 1998). Thus,
in this  article we  present a method  to compute  the one-point PDF  of the
aperture mass, both for the  quasilinear and strongly non-linear regimes. In
the case of the quasilinear  regime we can use rigorous perturbative methods
while in the highly non-linear regime we have to use a specific hierarchical
tree  model (which has  been seen  to agree  reasonably well  with numerical
simulations). Although the details of  the calculations are specific to each
case  we point  out the  general  pattern common  to both  regimes which  is
brought  about by  the projection  effects. In  particular, our  methods are
quite general and actually apply to any filters, though we are restricted to
axisymmetric  filters for  the  quasi-linear regime.   Our  results for  the
non-linear  regime, where  there  is not  such  a restriction,  can also  be
extended to multivariate statistics ($p-$point PDFs).
%The reason
%is that it is possible to relate the local mass aperture to the
%shear field only.

In Sect. 2  we recall the definitions of the  local convergence and aperture
mass and  how they  are related  to the cosmic  3D density  fluctuations. In
particular we present  the shape of the compensated filters  that we use for
the  explicit  computations we  present  in the  following.  In  Sect. 3  we
describe  the  relationship between  the  PDF  and  the cumulant  generating
function  of the  3D  density field  and we  show  how this  extends to  the
projected density. The details of the calculations are presented in Sect. 4,
for   the  quasilinear   theory,  and   in   Sect.  5   for  the   nonlinear
theory. Numerical results are presented in Sect. 6.

\section{The convergence and aperture mass fields}
\label{Definition}

In weak lensing observations, background galaxy deformations can be
used to reconstruct the local gravitational convergence field.
We recall here how the local convergence is related to the line-of-sight
cosmic density fluctuations. As a photon travels
from a distant source towards the observer its trajectory is perturbed
by density fluctuations close to the line-of-sight. This leads to an
apparent displacement of the source and to a distortion of the
image. In particular, the convergence $\kappa$ magnifies (or
de-magnifies) the source as the cross section of the beam is decreased
(or increased). One can show (Kaiser
1998) that the convergence along a given line-of-sight is,
\be
\kappa = \int_0^{\drad_s} \d\drad \; \wh(\drad,\drad_s) \; \delta(\drad)
\label{kappa}
\ee
when lens-lens couplings and departure from the Born approximation are
neglected (e.g., Bernardeau et al. 1997). This equation states that
the local convergence is obtained by an integral over the line-of-sight
of the local density contrast. The integration variable is the radial
distance, $\drad$, (and $\drad_s$
corresponds to the distance of the source) such that
\be
\d\drad = \frac{c\,\d z/H_0}{\sqrt{\Ol+(1-\Om-\Ol)(1+z)^2+\Om(1+z)^3}}
\label{chi}
\ee
while the angular distance $\De$ is defined by,
\be
\De(z) = \frac{c / H_0}{\sqrt{1-\Om-\Ol}} \sinh \left( \sqrt{1-\Om-\Ol} \;
\frac{H_0 \drad}{c} \right)
\label{De}
\ee
Then, the weight $\wh(\drad,\drad_s)$ used in (\ref{kappa}) is given by:
\be
\wh(\drad,\drad_s) = \frac{3 \Omega_m}{2} \; \frac{H_0^2}{c^2} \; \frac{\De(\drad)
\De(\drad_s-\drad)}{\De(\drad_s)} \; (1+z)
\label{w}
\ee
where $z$ corresponds to the radial distance $\drad$. Thus the
convergence $\kappa$ can be expressed in a very simple fashion as a
function of the density field. We can note from (\ref{kappa}) that
there is a minimum value $\kappamin(z_s)$ for the convergence of a
source located at redshift $z_s$, which corresponds to an ``empty''
beam between the source and the observer ($\delta=-1$ everywhere along
the line-of-sight): 
\be
\kappamin = -  \int_0^{\drad_s} \d\drad \; \wh(\drad,\drad_s)
\label{kappamin}
\ee
In practice, rather than the convergence $\kappa$ it can be more convenient
(Schneider 1996) to consider the aperture mass $\Map$. It corresponds to a
geometrical average of the local convergence with a window of vanishing
average,
\be
\Map =
\int\d^2\,\vartheta'\,U(\vartheta')\,\kappa(\vec\vartheta'-\vec\vartheta)  
\label{Map}
\ee
where $\kappa({\vec \vartheta})$ is the local convergence at the angular
position ${\vec \vartheta}$ and the window function $U$ is such that
\be
\int \d^2\vartheta \,U(\vartheta) = 0.
\label{IntU}
\ee
In this case, $\Map$ has the interesting property that it can be
expressed as a function of the tangential component $\gam_t$ of the
shear (Kaiser et al. 1994; Schneider 1996) so that it is not in principle
necessary to build local shear maps to get local aperture mass
maps. More precisely we can write, 
\be
\Map(\vartheta) = 
\int \d^2 \vartheta' \, Q(\vartheta') \, \gam_t({\vec\vartheta-\vec
\vartheta'}) 
\ee
with
\be
Q(\vartheta) = - U(\vartheta) + 
\frac{2}{\vartheta^2} \int \d\vartheta' \,\vartheta' \, U(\vartheta').
\label{Qzeta}
\ee
Nonetheless considering such a class of filters is interesting because
convergence maps are always reconstructed to a mass sheet degeneracy only.
Therefore, to some extent, any statistical quantities that can be measured
in convergence mass maps correspond to smoothed quantities 
with compensated filters.
In the following, for convenience rather than due to intrinsic
limitation of the method,
we consider filters that are defined on a compact support.

\subsection{Choice of filter}

\begin{figure}
{\epsfxsize=8.5cm {\epsfbox{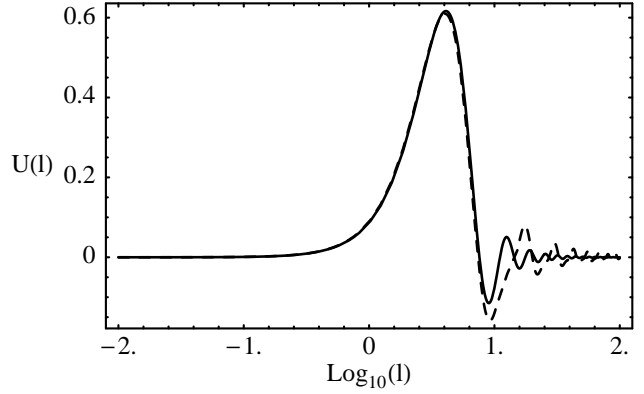}} }
\caption{Shape of the filter functions we use in Fourier space. The
solid line corresponds to the shape proposed by Schneider, Eq. (\ref{Ups}), 
multiplied by 1.459 and the dashed line corresponds to the compensated filter
we introduce in this paper, Eq. (\ref{Ubv}).}
\label{FilterShape}
\end{figure}

It is convenient to write the filter function in terms of reduce
variable, $\vartheta/\theta$ where $\theta$ is the filter 
scale,
\be
U(\vartheta) = \frac{ u(\vartheta/\theta) }{\theta^2}
\label{Uu}
\ee
so that the evolution with $\theta$ of the properties of $\Map$ only depends
on the  behavior of the  density field seen  on different scales  (while the
shape  and the normalization  of the  angular filter  $U(\vartheta)$ remains
constant).  In the  following  we  shall use  two  different filters,  which
satisfy  (\ref{Uu}).   One,  which  we  note $u_{S}$,  has  been  explicitly
proposed by Schneider (1996),
\be 
u_S(x)= \frac{9}{\pi} \; (1-x^2) \; \left( \frac{1}{3} - x^2 \right) \ \
{\rm  for}\ x<1  
\ee 
and  $u_S(x)=0$ otherwise.   The Fourier  form  of this
filter corresponds  to, \be W_{S}(l)=  2\pi\int_0^1 \d x  \; x \;  u_S(x) \;
J_0(l x) = {24\,J_4(l)\over l^2}.
\label{Ups} 
\ee 
The other,  which we note $u_{BV}$,
corresponds  to a  simpler compensated  filter that  can be  built  from two
concentric discs.
It is built from the  difference of the average convergence in a disc
$1/2$ and  the average convergence in a  disc unity:
\be
u_{BV}(x) = {4 H(2 x)\over \pi} - {H(x)\over \pi}
\label{uBV}
\ee
where $H(x)$ is the characteristic function of a disc unity.
In Fourier  space it is
simply related to the Fourier transform of a disc of radius unity,
\be
W_{1}(l)={2\,J_1(l)\over l},\label{W2D}
\ee
and reads,
\be 
W_{BV}(l)=W_1(l/2)-W_1(l)={4\,J_{1}(l/2)\over l}-{2\,J_1(l)\over l}.\label{Ubv} 
\ee 
The ratio  of the 2 radii has been chosen  so that the 2
filters are close  enough, as shown in Fig. 1.  Note that the normalizations
are somewhat arbitrarily. In the plot they have been chosen to give the same
amplitude for the aperture mass fluctuations in case of a power law spectrum
with index $n=-1.5$. This is obtained by multiplying the expression
(\ref{Ups}) by 1.459.
Fig. 1 shows actually that the two filters are very close to each other.  In
particular they have their maximum at the same $k$ scale, and except for the
large $k$ oscillations they exhibit a similar behavior.  In the following we
will take the freedom to use either one or the other for convenience.

\section{One-point PDF construction}

In this section we succinctly review the theory for the construction
of the one-point PDF statistical quantities. In particular we recall
the mathematical relationship between the one-point PDF and the
moment and cumulant generating functions.

\subsection{General formalism}

In general one can define  $\chi(\lambda)$ as the generating function of the
cumulants of a given local random variable $\delta$,   
\be
\chi(\lambda)=\sum_{p=1}^{\infty}  \mg \delta^p\md_c\,{\lambda^p\over  p!}  . 
\ee  
It is given through a Laplace transform of the one-point PDF
of the local density contrast $\delta$,
\be
e^{\chi(\lambda)}=\int_{-1}^{\infty}\d\delta\,e^{\lambda \delta}\,P(\delta).
\ee
For hierarchical models, that is when  
$\mg   \delta^p\md_c   \sim   \mg\delta^2\md^{p-1}$
it is convenient to define $\varphi(y)$ as   
\be
\varphi(y)=-\sum_p{\mg\delta^p\md_c\over\mg\delta^2\md^{p-1}}\,{(-y)^p\over
p!}= - \mg\delta^2\md\,\chi(-y/\mg\delta^2\md). 
\label{phidef} 
\ee 
Then the one-point PDF of $\delta$ is then given by the inverse Laplace
transform (see Balian \& Schaeffer 1989), 
\be  
P(\delta)\d\delta=\d\delta\int{\d
y\over  2\pi\ii\sigma^2} \exp\left[-{\varphi(y)\over \sigma^2}+{y\delta\over
\sigma^2}\right],  
\label{Pdelta}
\ee  of   the  moment   generating  function.  Here   $\sigma=\lag  \delta^2
\rag^{1/2}$ is the r.m.s. density fluctuation.

\subsection{The projection effects}

The relation (\ref{kappa})  states that the local convergence  can be viewed
as  the superposition  of independent  layers  of cosmic  matter field.  The
direct  calculation of  the one-point  PDF of  such a  sum would  involve an
infinite number  of convolution products  which makes it intractable.  It is
more convenient  to consider the cumulant generating  functions which simply
add when different layers  are superposed (because Laplace transforms change
convolutions into ordinary products).

The projections effects for  statistical properties of the local convergence
have  already been  considered in  previous papers.   It has  been  shown in
particular how  the moments of the  projected density can be  related to the
ones  of the  3D  field in  both  hierarchical models  corresponding to  the
non-linear regime (T\'oth, Holl\'osi and Szalay 1989) and in the quasilinear
regime (Bernardeau 1995).

More recently it has been shown (Valageas 2000a,b; Munshi \& Jain 1999a) that
these results could be extended to the full PDF of the projected density. In
particular the cumulant generating function  of the projected density can be
obtained by  a simple  line-of-sight average of  the 3D  cumulant generating
function.

It is convenient to define the normalized projected density contrast
$\deltap$ by:
\be
\deltap= {\kappa\over\vert\kappa_{\rm min}\vert} =
\int\d\drad\,F(\drad)\,\delta(\drad),
\label{defdeltap}
\ee
where $F(\drad)$ is the selection function for the
projection effects as a function of the radial distance $\drad$:
\be
F(\drad)={\wh(\drad,\drad_s)\over \vert\kappa_{\rm min}\vert} .
\ee
When filtering effects are included we have,
\be
\deltapt=\int \d^2 \vartheta' \; w_{\theta}(\vartheta') \;
\deltap(\vec\vartheta'-\vec\vartheta)
\label{defdeltapt}
\ee
where $w_{\theta}$ is a given window function at scale $\theta$. A particular
case is provided by the normalized aperture mass $\Maph$,
\be
\Maph = \frac{\Map}{|\kappamin|} = \int \d^2 \vartheta' \; U(\vartheta') \;
\deltap(\vec\vartheta'-\vec\vartheta).
\label{defMaph}
\ee

The cumulants of the projected density can be related to those
of the 3D density fields. Formally they correspond to the ones
of the field when it is filtered by a conical shape window. Thus, from
(\ref{defdeltap}) we obtain:
\ba
\lag \deltap ({\vec \vartheta}_1) \dots \deltap({\vec
\vartheta}_p) \rag_c &=& 
\int_0^{\drad_s} \prod_{i=1}^{p} \d\drad_i \; F(\drad_i) \nonumber\\ 
&&\hspace{-2cm} \times \lag \delta(\drad_1, \De_1 {\vec
\vartheta}_1) \dots \delta(\drad_p, \De_p {\vec \vartheta}_p) \rag_c  . 
\label{cum1}
\ea
The computation of such quantities can be made in the small angle
approximation. Such approximation is valid when the transverse
distances $\De\vert\vec\vartheta_i-\vec\vartheta\vert$ are much
smaller than the radial distances $\drad$. In this case the integral
(\ref{cum1}) is dominated by configurations where
$\drad_i-\drad_j\sim \De_i\vert\vec\vartheta_i-\vec\vartheta_j\vert
\sim \De_j\vert\vec\vartheta_i-\vec\vartheta_j\vert$. It permits to make
the change of variables $\drad_i\to                r_i$                with
$\drad_i=\drad_1+r_i\De_1$. Then,
since the correlation length (beyond which the many-body correlation
functions are 
negligible) is much smaller than the Hubble scale $c/H(z)$ (where $H(z)$ is the
Hubble constant at redshift $z$) the integral over $r_i$
converges over a small distance of the order of
$\vert\vec\vartheta_i-\vec\vartheta_1\vert$ and the expression
(\ref{cum1}) can be simplified in,
\begin{eqnarray}
\lag \deltap  ({\vec \vartheta}_1) \dots  \deltap({\vec \vartheta}_p) \rag_c
&=&    \int_0^{\drad_s}    \De_1^{p-1}\d\drad_1    \;   F_1^p    \nonumber\\
&&\hspace{-4.5cm}     \times    \int_{-\infty}^{\infty}    \prod_{i=2}^{p}\;
\d   r_i   \;   \lag  \delta(\drad_1,   \De_1   {\vec
\vartheta}_1) \dots \delta(\drad_1+r_i\De_1, \De_1 {\vec \vartheta}_p) \rag_c .
\label{cum2}
\end{eqnarray}

Taking filtering effects into account leads to,
\ba
\mg\delta_{{\rm proj.},\theta}^p\md_c &=&
\int_0^{\drad_s} \De_1^{p-1}\d\drad_1 \; F_1^p \; \int \prod_{i=1}^{p}
{\d^2 \vartheta_i}w_{\theta}(\vartheta_i)  \nonumber\\ 
&&\hspace{-2cm} \times \int_{-\infty}^{\infty}    \prod_{i=2}^{p}
\d r_i \; 
\lag \delta(\drad_1, \De_1 {\vec
\vartheta}_1) \dots \delta(\drad_1+r_i\De_1, \De_1 {\vec \vartheta}_p) \rag_c.
\label{cum2bis}
\ea
Thus the projection effects reduce to
\be
\mg\delta_{{\rm proj.},\theta}^p\md_c=
\int\d\drad\,F^p(\drad)\,\mg\delta^p_{\De\theta,\,{\rm cyl.}}\md_c\,
L^{p-1},
\label{proj1}
\ee
where $\delta^p_{\De\theta,\,{\rm cyl.}}$ is the filtered 3D density
with a cylindrical filter of transverse size $\De \, \theta$ and depth
$L$ (which goes to infinity in (\ref{cum2})).

In particular this result gives the expression of the
variance of the filtered projected density contrast,
\be
\mg\deltapt^2\md=\int_0^{\drad_s}\d\drad\,F^2(\drad)\,
\mg\delta^2_{\De\theta,\,{\rm cyl.}}\md_c\,L
\ee
This expression can be re-expressed in terms of the power spectrum
$P(k)$ (nonlinear power spectrum), defined in this paper with
\be
\mg\delta(\vx_1)\delta(\vx_2)\md=\int{\d^3\vk\over (2\pi)^3}
P(k)\,e^{\ii\vk(\vx_1-\vx_2)}.
\ee
Then
\ba
\mg\delta^2_{\De\theta,\,{\rm cyl.}}\md_c&=&
\int_0^{\infty}{\d\kpar\over 2\pi}\,\int{\d^2\vk_{\perp}\over(2\pi)^2}\,
P(k)\,\nonumber\\
&&\times\left[{2\,\sin(\kpar\,L)\over
\kpar\,L}\right]^2\,W^2(\De\theta k_{\perp})  
\ea
where $\vk_{\perp}$ is the component of $\vk$ orthogonal to the radial
direction and $\kpar$ is the component along the line-of-sight.
In the previous integral, $\kpar\sim 1/L$ and
$k_{\perp}\sim1/(\De\theta)$, so that 
when $L$ is large $\kpar$ is negligible compared to $k_{\perp}$ which
leads to,
\be
\mg\delta^2_{\De\theta,\,{\rm cyl.}}\md={1\over
L}\int{\d^2\vk_{\perp}\over(2\pi)^2}\,P(k_{\perp}) 
\,W^2(\De\theta\vk_{\perp}) 
\label{varcyl}
\ee
where $W$ is the Fourier shape of the 2D window function. This
relation holds for the filtered projected density contrast as well as
the aperture mass, for which $W$ in (\ref{varcyl}) is to be replaced
by $W_S$ or $W_{BV}$.

The formal expression for the higher order moments can be
simplified by taking advantage of the so-called scaling
laws for the correlation functions. It is in particular
natural to assume that,
\be
\mg\delta^p\md_c \propto \mg\delta^2\md^{p-1} \label{scalings}
\ee
with a coefficient of proportionality, $S_p$, that depends on both
the power spectrum  and filter shapes, but not on the
power spectrum normalization. For power law spectrum it implies
in particular that these coefficients do not depend
on the filtering scale. In the coming sections we present
in more details the origin of this scaling relation. It allows to define
\be
\varphi(y)\equiv-\sum_p S_p\,{(-y)^p\over p!}=-
\mg\delta^2\md\,\chi(-y/\mg\delta^2\md).
\ee
The equation (\ref{proj1}) then relates the cumulant generating
function $\varphi(y)$ for the projected density to the one corresponding
to cylindrical filtering effects,
\be
\varphip(y)=\int {\d \drad\over \psi_{\theta}(\drad)}\,
\varphic[y\,F(\drad)\psi_{\theta}(\drad)]\label{phiproj}
\ee
with
\be
\psi_{\theta}(\drad)={\mg\delta^2_{\De\theta,\,{\rm cyl.}}\md\over 
\mg\deltapt^2\md} \, L
\ee
which can be rewritten in terms of the matter fluctuation
power spectrum,
\be
\psi_{\theta}(\drad)={\int\d^2\vk\,P(k,z)\,W^2(k\,\De\,\theta)\over
\int\d\drad'\,F^2(\drad')\,\int\d^2\vk\,P(k,z')\,W^2(k\,\De'\,\theta)}.
\ee
In this expression we have explicitly written the redshift dependence
of the power spectrum.
In case of a power law spectrum,
\be
P(k,z)=P_0(z) \,\left({k\over k_0}\right)^n
\ee
it takes a much simpler form given by,
\be
\psi_{\theta}(\drad)={P_0(z)\,\De^{-n-2}\over 
\int\d\drad'\,F^2(\drad')\,P_0(z')\,\De'^{-n-2}}.
\ee
The result (\ref{phiproj}) 
is the cornerstone of the calculations we present. It
allows to relate the cumulant generating function of projected
quantities to the ones computed in much simpler geometries.

The difficulty then resides in the computation of $\varphic$ in the regimes
we are interested in. Two limit cases are actually accessible to exact
calculations. First, one is the quasilinear regime where one can take advantage of the
special properties of the perturbative expansion in Lagrangian space to
build the cumulant generating function in Eulerian space. These kinds of
properties had been used previously for 3D or 2D top-hat filters only.
We show here how this method can be extended to the filter (\ref{Ubv}).
Second, in the strongly nonlinear regime one can also do exact numerical calculations
when one is assuming the high order correlation functions to follow a
tree model. The derivations corresponding to these regimes are
presented in the next two sections.

\section{The quasi-linear regime}

The reason rigorous calculations can be carried on
in this regime is that the cumulant
generating function for a cylindrical shape is exactly the one corresponding
to the 2D dynamics (Bernardeau 1995). In other words, in this case
\be
\varphic^{\rm quasilinear}(y)=\varphi^{\rm quasilinear}_{2D}(y)
\ee
at leading order.
In this regime the problem then reduces to the computation of cumulant
generating functions for compensated filters for the 2D dynamics.
The latter calculation is presented in the following paragraphs. This
is a long and technical calculation that leads to the formulae
(\ref{eqphieff}-\ref{eqzetaeff}).

The calculations we present follow what has been done for the top-hat window
filters, with  the complication  introduced by the  use of two  such filters
instead of one to build the compensated filter. 

\subsection{2D statistics in Lagrangian space}

The generating function for the compensated filter (\ref{uBV}) can be built
from the generating  function of the  joint density PDF  for two
concentric cells of  different radius.  This quantity will be obtained in
Eulerian space from the one in Lagrangian space through 
a Lagrangian-Eulerian mapping. 
More precisely  we consider  the joint  PDF of  two reduced volumes
defined as the comoving volume $V_c(t)$ occupied by some matter
expressed in units of the volume $V$ it occupied initially, 
\be v={V_c(t)\over V}.  
\ee 
The Eulerian overdensity of this matter region will then be given by the inverse
of $v$, 
\be
\rho=1/v.
\ee
The calculation will be made in two steps. First we present the
derivation of the cumulant generating function in Lagrangian space,
then the mapping from Lagrangian to Eulerian space.

In a Lagrangian description $v$ corresponds to the Jacobian of the
transform from the initial coordinates $\vq$ in Lagrangian
space to the ones in real space $\vx$,
\be
v=J(\vq)=\left\vert{\partial\vx\over\partial\vq}\right\vert.
\ee
The construction of the volume PDF is then based 
on the geometrical properties of the Jacobian
perturbative expansion. Its expansion  with respect to the
initial density fluctuations (in the rest of this subsection
we consider 2D dynamics) reads
\be
J(\vq)=1+J^{(1)}(\vq)+\dots
\ee
Each term of this expansion can be written in terms of the
initial Fourier modes of the linear density field,
\ba
J^{(p)}&=&\int{\d^2\vk_1\over 2\pi}\dots{\d^2\vk_p\over 2\pi}
D_+^p(t)\,\delta(\vk_1)\dots\delta(\vk_p)\times\nonumber\\
&&\exp[\ii(\vk_1+\dots+\vk_p)\vq]\,J_p(\vk_1,\dots,\vk_p).
\ea
where $D_+(t)$ describes the time dependence of the linear growing
mode. The central issue is the way the geometrical
kernel $J_p(\vk_1,\dots,\vk_p)$ behaves when geometrical 
effects are taken into account. At leading order in perturbation
theory (that is when only ``tree order'' terms are taken into account)
that amounts to compute terms of the
form
\be
\int \d\alpha_1\dots\d\alpha_p\,
J_p(\vk_1,\dots,\vk_p)\,W\vert\vk_1+\dots+\vk_p\vert.
\ee
where  $\alpha_i$ is the angle of the $i^{\rm th}$ wave vector. There exists a central property, valid for top-hat filters only, which states that (Bernardeau 1995),
\ba
&&\int \d\alpha_1\dots\d\alpha_p\,
J_p(\vk_1,\dots,\vk_p)\,W_1\vert\vk_1+\dots+\vk_p\vert=\\
&&\ \ \ W_1(k_1)\dots W_1(k_p)\,\int \d\alpha_1\dots\d\alpha_p\,
J_p(\vk_1,\dots,\vk_p),\nonumber
\ea
where $W_1$ is defined in (\ref{W2D}).
This result extends the one obtained in Bernardeau (1994) for the 3D
dynamics.

At leading order in Perturbation Theory any cumulant of the form
$\mg v_1^p\,v_2^q\md_c$ involves only products of such quantities.
The sort of commutation rule given in the previous 
equation implies that these cumulants can be computed without explicitly 
taking into account the filtering effects.
It means that any cumulant of the form $\mg v_1^p\,v_2^q\md_c$
can be built with a {\em tree shape construction} with two different
kinds of end points. Formally the generating function of such cumulants 
\be
\chi(\lambda_1,\lambda_2)=\sum_{p,q}{\lambda_1^p\over p!}{\lambda_2^q\over q!}
\mg v_1^p\,v_2^q\md_c
\ee
reads,
\ba
\chi(\lambda_1,\lambda_2)&=&\lambda_1\zeta_J(\tau_1) + \lambda_2\zeta_J(\tau_2)
-\label{chirel2}\\
&&\ \ \ \ {\lambda_1\over 2}\tau_1\,\zeta_J'(\tau_1)
-{\lambda_2\over 2}\tau_2\,\zeta_J'(\tau_2)\nonumber\\
\tau_1&=&\lambda_1\xib_{11}\zeta_J'(\tau_1)+\lambda_2\xib_{12}\zeta_J'(\tau_2)\\
\tau_2&=&\lambda_1\xib_{12}\zeta_J'(\tau_1)+\lambda_2\xib_{22}\zeta_J'(\tau_2), 
\ea
where $\xib_{ij}$ are the second moment of the linear density contrasts
between two cells of fixed Lagrangian radii $\theta_i$ and $\theta_j$,
\be
\xib_{ij}\equiv\xib_{ij}(\theta_i,\theta_j)=
\int{\d^2\vk\over (2\pi)^2}\,P(k)\,W(k\theta_i)\,W(k\theta_j).\label{xibij}
\ee
They are, for the Lagrangian variables $v_i$, fixed parameters
that depend only on the power spectrum shape (and on the set of cell
radii chosen at the beginning).
The function $\zeta_J$ is the generating function of the
angular averages of $J_p(\vk_1,\dots,\vk_p)$,
\ba
\zeta_J(\tau)&=&
\sum_{p=0}^{\infty}\,j_p\,{\tau^p\over p!},\ \ \ \ j_0=1\\ 
j_p&=&{1\over (2\pi)^p}\int \d\alpha_1\dots\d\alpha_p\,
J_p(\vk_1,\dots,\vk_p).
\ea
Note that $\zeta(\tau)-1=1/\zeta_J(\tau)-1$ describes the density
contrast of a spherical density fluctuation of linear over-density $\tau$ for the
2D dynamics.
The exact form of $\zeta_J$ is therefore known for any cosmological model.

The relation (\ref{chirel2}) 
can actually be generalized to an arbitrary number of cells
in a straightforward way,
\ba
\chi(\{\lambda_i\})&=&\sum_{i=1}^n\lambda_i\zeta_J(\tau_i)
-{\lambda_i\over 2}\tau_i\,\zeta_J'(\tau_i)\\
\tau_i&=&\sum_{j=1}^n\lambda_j\,\xib_{ij}\,\zeta_J'(\tau_j).\label{tauirel}
\ea
The latter relation can be rewritten in an equivalent way as
\be
\sum_{j=1}^n C_{ij}\,\tau_j=\lambda_i\,\zeta'_J(\tau_i)
\ee
where $C_{ij}$ is the inverse matrix to $\xib_{ij}$.
The generating function $\chi$ can then be written,
\be
\chi(\{\lambda_i\})=\sum_{i=1}^n\lambda_i\,\zeta_J(\tau_i)-
{1\over 2}\sum_{i,j=1}^n C_{ij}\,\tau_i\,\tau_j.
\ee
It gives the generating function of the reduced volume generating
function of an arbitrary number of concentric cells. Note that 
however the known geometrical properties of the Lagrangian expansion
terms do not allow to extend these results to non-concentric cells. Thus, our 
method actually apply to any filter which is axisymmetric (in general one would need an infinite number of cells but in practice numerical discretization always leads to a finite number of concentric shells). 
It is worth noting that the relation (\ref{tauirel}) gives,
\be
\sum_j\lambda_j\zeta_J'(\tau_j){\d\tau_j\over\d\lambda_i}-
\sum_j\tau_j\zeta_J''(\tau_j){\d\tau_j\over\d\lambda_i}=\tau_i\zeta'_J(\tau_i)
\ee
which in turns leads to,
\be
{\partial\chi(\{\lambda_i\})\over\partial\lambda_i}=\zeta(\tau_i).
\label{dchiprop}
\ee

\subsection{Saddle point approximation and leading order
cumulant generating function}

In this subsection we explicit the formal relationship
between the generating function computed at leading order
and the shape of the  multivalued density 
probability distribution function.

The joint PDF is formally given by (this is an extension of Eq. \ref{Pdelta})
\ba
P(v_1,\dots,v_n)&=&\int{\d\lambda_1\over 2\pi\ii}\dots
{\d\lambda_n\over 2\pi\ii}\\
&&\times\exp\left[\chi(\{\lambda_i\})-\sum_{i=1}^n\lambda_i\,v_i\right].\nonumber
\ea
In case of a small variance, the expression of the joint
density is obtained through the saddle point approximation.
The saddle point conditions read,
\be
{\partial \chi\over \partial \lambda_i}=v_i,
\ee
which gives implicitly the values of $\lambda_i$ at the saddle
point position in terms of $v_i$. Taking advantage of the property
(\ref{dchiprop}), one gets,
\be
\zeta_J(\tau_i)=v_i.\label{nlmapping}
\ee
It implies that with the saddle point position the expression of the
joint PDF is (not taking into account prefactors),
\be
P_{\rm Lag.}
(v_1,\dots,v_n)\sim\exp\left[-{1\over2}\sum_{ij}C_{ij}\,\tau_i\,\tau_j\right]
\ee
with the mapping (\ref{nlmapping}). This is exactly
what one would expect for Gaussian initial conditions, $\tau_i$
being the linearly extrapolated density contrasts at the chosen scales.

\subsection{Lagrangian-Eulerian mapping}

To relate Lagrangian and Eulerian space, one uses the same trick as
in Bernardeau (1994), that is,
\ba
&&P_{\rm Lag.}\left(v_1>{1\over \rho_{01}},\dots,v_n>{1\over
\rho_{0n}}\right)=\nonumber\\
&&\ \ \ \ \ \ \ \ \ \ \ \ \ \ \ \ \ \ 
P_{\rm Eul.}(\rho_{1}<\rho_{01},\dots,\rho_n<\rho_{0n}).
\ea
The leading order cumulant generating function can then be obtained by an
identification of the exponential term when one uses the saddle
point approximation. The variables are however now changed in $\rho_i$
which are related to $\tau_i$ with
\be
\zeta(\tau_i)=\rho_i, \ \ \zeta(\tau_i)=1/\zeta_J(\tau_i).
\ee
Moreover the variables $\rho_i$ enter also the expression
of the cell correlation coefficients $\xib_{ij}$ since they are
in Eq. (\ref{xibij}) computed for a fixed mass scale and not for a
fixed Eulerian space radius. As a result, the coefficient $C_{ij}$
expressed in terms of $\xib_{ij}$ should be understood
as function of the variable $\rho_i$ through
\ba
\xib_{ii}&=&\xib\left(\rho_i^{1/2}\,\theta_i\right)\\
\xib_{ij}&=&\xib\left(\rho_i^{1/2}\,\theta_i,\rho_j^{1/2}\,\theta_j\right)
\ea
where $\theta_i$  are all  kept fixed. 

The cumulant generating function in Eulerian space is obtained
also with a saddle point approximation in the computation of
\ba
\exp\left[\chi_{\rm Eul}(\{\lambda_i\})\right]&=&
\int\d\rho_1\dots\d\rho_n
\,P_{\rm Eul}(\rho_1,\dots,\rho_n)\nonumber\\
&&\ \ \ \ \ \ \ \times\exp\left(\sum_i\lambda_i\rho_i\right)
\ea
which leads to,
\be
\chi_{\rm Eul}(\{\lambda_i\})=
\sum_{i=1}^n \lambda_i\rho_i-{1\over 2}\sum_{i,j=1}^n C_{ij}\,\tau_i\,\tau_j\label{chieul}
\ee
with the stationary conditions,
\be
{1\over 2}{\partial \over \partial \rho_i}
\sum_{i,j=1}^n C_{ij}\,\tau_i\,\tau_j=\lambda_i \label{statcond}
\ee
where the partial derivatives should then be understood for
fixed radius $\theta_i$ and $\lambda_i$. The relation (\ref{chieul}),
together with the conditions (\ref{statcond}) gives the formal expression
of the cumulant generating function in Eulerian space.

\begin{figure}
{\epsfxsize=8cm {\epsfbox{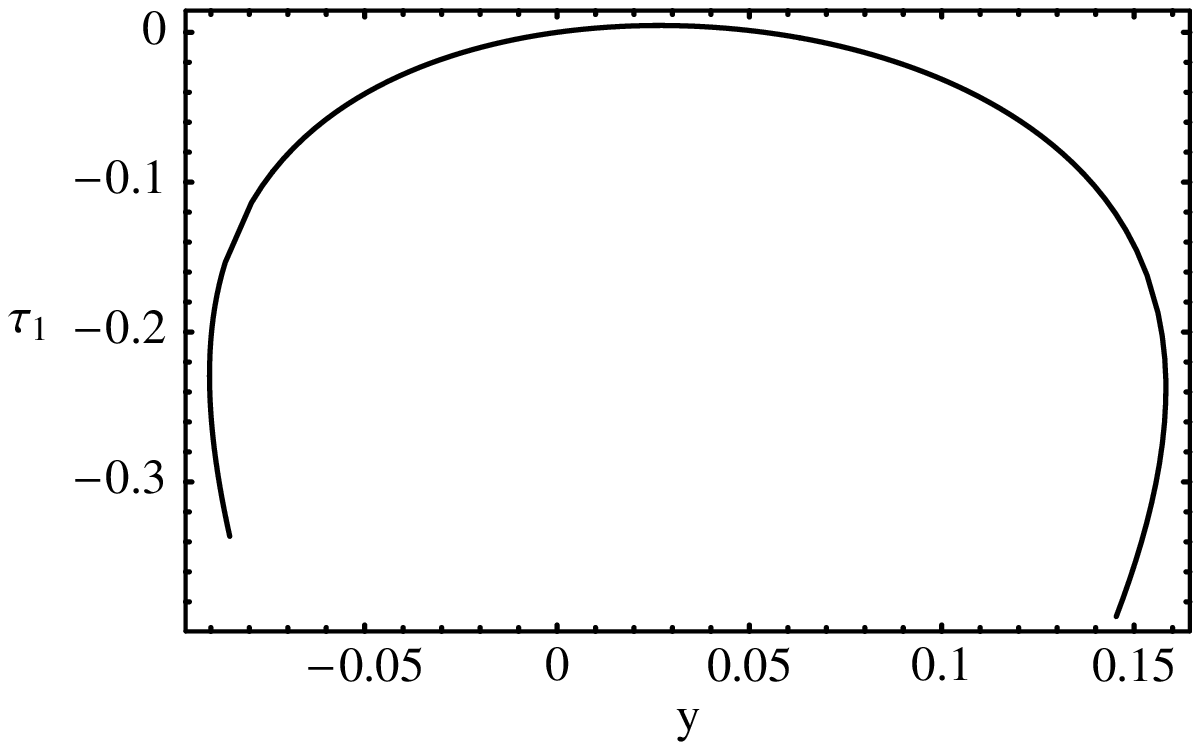}} }
{\epsfxsize=8cm {\epsfbox{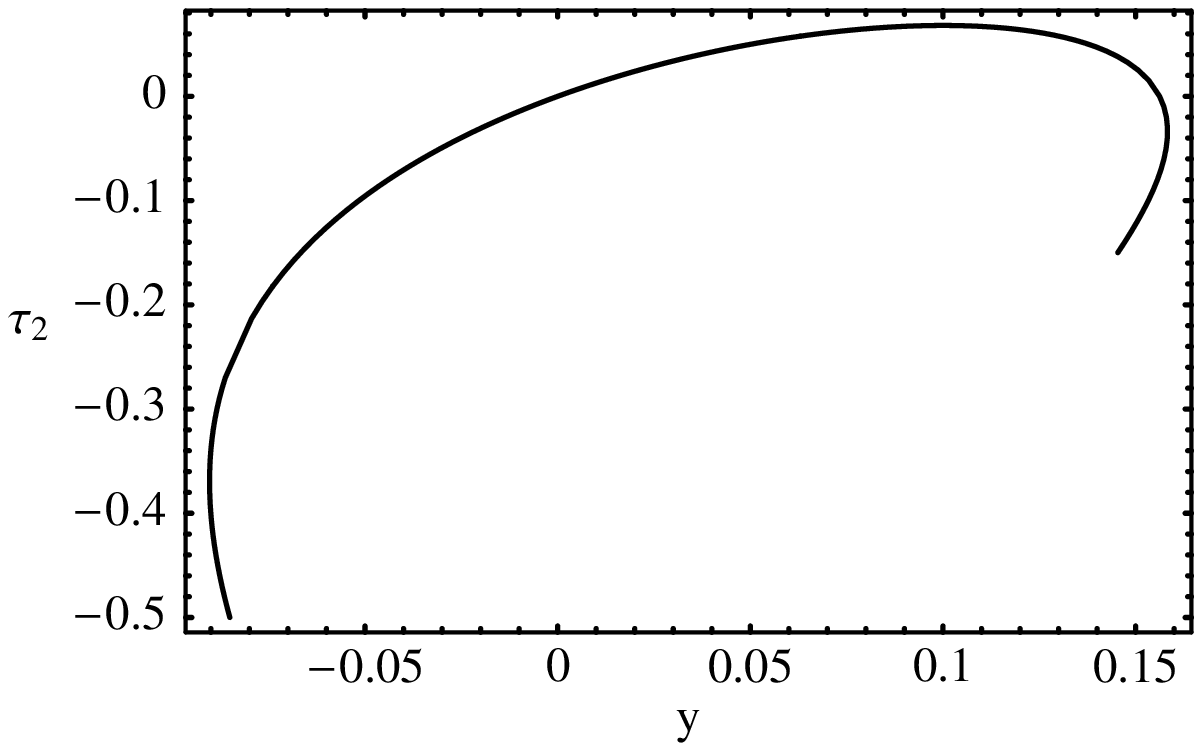}} }
{\epsfxsize=8cm {\epsfbox{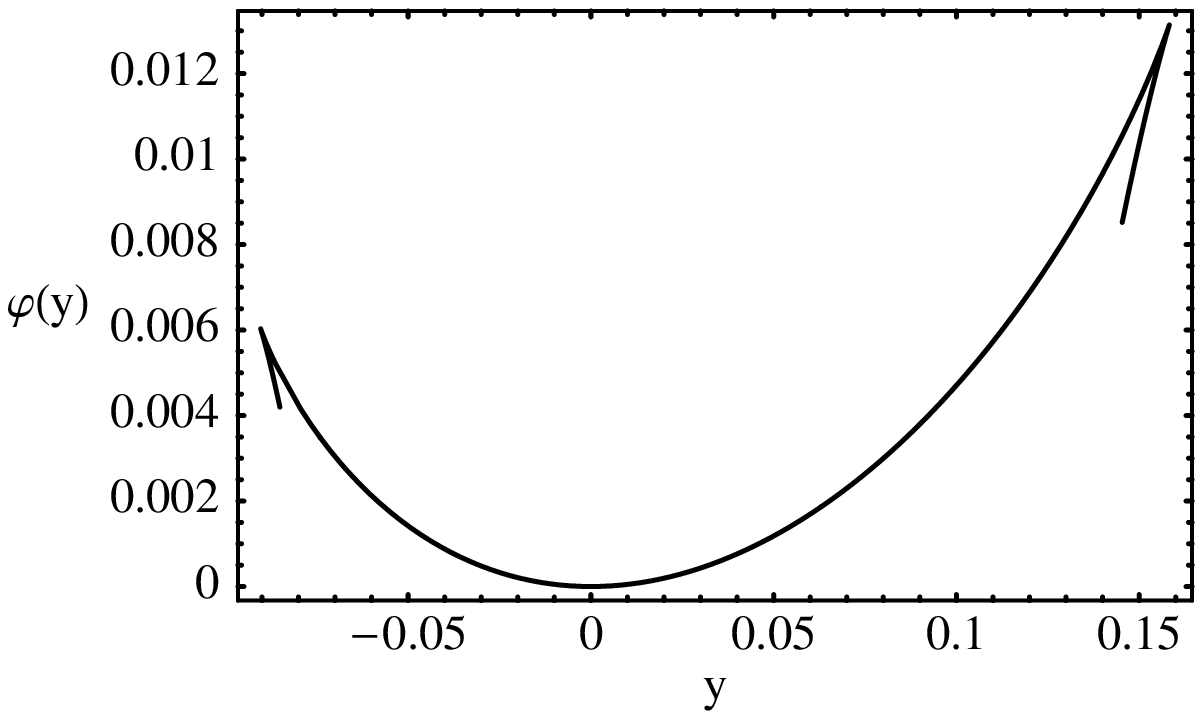}} }
\caption{Shape of the functions $\tau_1(y)$ and $\tau_2(y)$
for the two-cell compensated window function ($y\equiv -\lambda\sigma^2$).
The computations have been done for a power law spectrum for $n=-1.5$.
The solution for $\tau$ has been extended beyond the singularity to
show explicitly that the singularities are due to double solution in $y$.} 
\label{PTresults}
\end{figure}

The case we are interested in,
\be
\Map=\rho_1-\rho_2
\ee
if
\be
\theta_1=1/2,\ \ \theta_2=1
\ee
corresponds to 2 cells, if $\Map$ is built with the filter (\ref{Ubv}).
Then the generating function for $\Map$ is obtained with a peculiar
choice for $\lambda_i$,
\be
\lambda=\lambda_1=-\lambda_2.
\ee
It is actually convenient to define,
\be
y=-\lambda\,\sigma^2
\ee
and
\be
\varphi_{2D}(y)=\varphic(y)=-\chi(\lambda)\,\sigma^2
\ee
where $\sigma^2$ is defined by 
\be
\sigma^2=\mg\Map^2\md= \xib_{11}+\xib_{22}-2\xib_{12}.
\ee
With this choice of variable, $\varphi(y)$
does not depend formally on the variance but only on $y$.
To be more specific one has finally,
\ba
\varphic(y)&=&y\zeta(\tau_1)-y\zeta(\tau_2)+\mF(\tau_1,\tau_2) \label{phicylql1}\\
\mF(\tau_1,\tau_2)&=&{\sigma^2\over 2[1-r(\tau_1,\tau_2)]}\nonumber\\
&&
\hspace{-0.5cm} \times\left[{\tau_1^2\over\xib_{11}(\tau_1)}+{\tau_2^2\over\xib_{22}(\tau_2)}
-{2r(\tau_1,\tau_2)
\tau_1\tau_2\over\sqrt{\xib_{11}(\tau_1)\xib_{22}(\tau_2)}}
\right],
\ea
where $r=\xib_{12}/\sqrt{\xib_{11}\xib_{22}}$, $\xib_{11}$ and 
$\xib_{22}$ are considered as function of $\tau_1$ and $\tau_2$
through the variables $\rho_1$ and $\rho_2$. the saddle point
conditions then read,
\ba
{\partial \mF\over\partial \tau_1}&=&-y\,\zeta'(\tau_1)\\
{\partial \mF\over\partial \tau_2}&=&y\,\zeta'(\tau_2). \label{phicylql2}
\ea

In this case the function $\chi(\lambda)$ can be numerically
calculated. We have restricted our calculations to the case where
the power spectrum follows a power law behavior with index $n=-1.5$.
To do the numerical computations we also use a simplified expression
for the 2D spherical collapse dynamics,
\be
\zeta(\tau)=(1+\tau/\ka)^{-\ka}
\ee
with
\be
\ka={\sqrt{13}-1\over 2}\approx 1.30.
\ee
The resulting function $\varphic(y)$ is shown in Fig. \ref{PTresults}, 
together with the functions $\tau_1(y)$ and $\tau_2(y)$. 

\subsection{Properties of the cumulant generating function}

These figures clearly show that the function $\varphi(y)$
has two singularities on the real axis. This is to be compared 
to what  is encountered for counts-in-cells statistics where
only one singular point is expected.
The numerical resolutions have been extended slightly beyond the
singularities to show that they are
due to the resolution of the implicit equations in $\tau$ that 
have multiple solution in $y$.
As a result the generic behavior near any of such singularity is
\ba
\tau(y) &\sim& t_s\,(y-y_s)^{1/2},\\
\varphic(y)&\sim& a_s\,(y-y_s)^{3/2}. \label{omsL}
\ea
This behavior directly induces exponential cutoffs for the shape
of the density PDF (see Balian \& Schaeffer 1989, Bernardeau \& Schaeffer
1992). In case of compensated filter, the fact that we obtain 2
singularities, induces two exponential cut-offs on both side
of the PDF as it appears clearly on the results presented in
Sect. \ref{Statistics of the aperture mass} (see also Valageas 2000c).

\begin{figure}
{\epsfxsize=8cm {\epsfbox{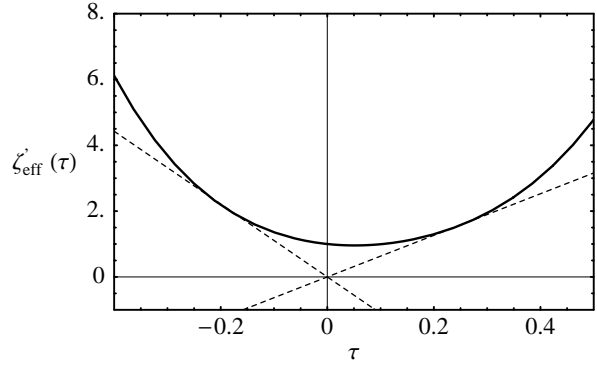}} }
\caption{Geometrical representation of the equation (\ref{eqtaueff}). The function $\tau(y)$ is given by the intersection of the line $-\tau/y$ with the curve $\zeta'_{\rm eff.}(\tau)$. Singularities appear when both curves are tangential, as shown in the figure.} 
\label{zetaprimedetau}
\end{figure}

In a phenomenological way the function $\varphic(y)$ can be described
by an effective vertex generating function $\zeta_{\rm eff.}(\tau)$  so
that,
\ba
\varphic(y)&=&y\,\zeta_{\rm eff.}
(\tau)-{1\over 2}y\,\tau\,\zeta'_{\rm eff.}(\tau)\label{eqphieff}\\
\tau&=&-y\,\zeta'_{\rm eff.}(\tau).\label{eqtaueff}
\ea
Numerically the effective vertex generating function is well
described by a fifth order polynomial,
\ba
\zeta_{\rm eff.}(\tau)&=&
-\tau + 0.843179\,{{\tau}^2} - 5.7511\,{{\tau}^3} + 3.3669\,{{\tau}^4}
-\nonumber\\
&&\ \ \ \ \ \ \ \ \ \ \ 6.3852\,{{\tau}^5}
\label{eqzetaeff}
\ea 
which is regular, the  expected singular behavior for $\varphi(y)$ being
induced by Eq. (\ref{eqtaueff})  (see Fig. \ref{zetaprimedetau}).  Note that
$\zeta_{\rm eff.}(\tau)$ can be viewed as a generating function of effective
vertices.  It  implies  that  for  instance  the  skewness  of  the  2D  (or
equivalently  cylindrical)  compensated  filtered  density is  
\be  
S_3^{\rm cyl.}= 3 \times  0.843179\approx 2.52 
\ee 
for such power  law spectrum shape. It is however important  to have in mind
that the shape of $\zeta_{\rm eff.}(\tau)$ as well as the skewness depend on
the window  function {\em normalization}.  The calculations have  been given
here  for the  $u_{BV}$  filter defined  in  Eq.  (\ref{Ubv}).  For the  $u_S$
filter, Eq. (\ref{Ups}), $S_3^{\rm cyl.}$ for instance would have been about
1.459 times larger, because of the normalization discrepancy between the two
filters.

The  skewness  of $\Map$  is  then  related to  this  one  through a  simple
projection factor,
\be 
S_3^{\rm  proj.}=S_3^{\rm cyl.}\frac{ \int \d\drad \,
F^3 \, \left[  P_0(z) \, \De^{-(n+2)} \right]^{2} }{  \left[ \int \d\drad \,
F^2 \,  P_0(z) \,  \De^{-(n+2)} \right]^{2}}.  
\ee  
This latter  relation is
actually valid in both the quasilinear and the nonlinear regime.

\section{The nonlinear regime}

In the nonlinear regime, there exist no derivations from first
principles of the behavior of the high order correlation functions
of the cosmic density field. However, hints of its behavior
can be found. The {\em stable clustering ansatz} gives indication on
the expected amplitude of the high order correlation functions and how
they scale with the two-point one. The hierarchical tree model, and
more specifically the {\em minimal tree model}, allows to build
a coherent set of high-order correlation functions.

\subsection{The stable-clustering ansatz}
\label{Stable-clustering ansatz}

The stable clustering ansatz (Peebles 1980) simply states that the high order
correlation functions should compensate, in virialized objects,
the expansion of the Universe. It gives not only the growth 
factor of the two-point correlation, but also a scaling relation 
between the high-order correlation functions.

Expressed in terms of the coefficient $S_p$ -hence of the generating
function $\varphi(y)$ defined in (\ref{phidef})-  it means that
they are independent of time and scale.
As a consequence,  the knowledge of the evolution of the
power-spectrum $P(k)$, or of the two-point correlation function
$\xia$, is sufficient to obtain the full PDF of the local 
density contrast. This property has been
checked in numerical simulations by several authors (Valageas et
al. 2000; Bouchet et al. 1991; Colombi et al. 1997; Munshi et
al. 1999). In particular, the statistics of the counts-in-cells
measured in numerical simulations provide an estimate of the
generating function $\varphi(y)$ for 3D top-hat filters.

More precisely, in the highly
non-linear regime one considers the variable $x$ defined by: 
\be
x = \frac{1+\delta_R}{\xia} . 
\label{defx}
\ee
Then, using (\ref{Pdelta}), for sufficiently ``large'' density
contrasts the PDF $P(\delta_R)$ can be written as (Balian \&
Schaeffer 1989): 
\be
P(\delta_R) = \frac{1}{\xia^{2}}\,h(x)
\label{Phx}
\ee
when 
\be
\xia \gg 1, \,(1+\delta_R) \gg \xia^{-\om/(1-\om)}
\label{Phxcond}
\ee
where the scaling function $h(x)$ is the inverse Laplace transform of
$\varphi(y)$: 
\be
h(x) =  -\inta \frac{dy}{2 \pi \ii}\,e^{xy}\, \varphi(y).
\label{hphi}
\ee
In (\ref{Phxcond}) the exponent $\om$ comes from the behavior of
$\varphi(y)$ at large $y$. Indeed, from very general considerations
(Balian \& Schaeffer 1989) one expects the function $\varphi(y)$
defined in (\ref{phidef}) to behave for 3D top-hat filtering 
as a power-law for large $y$: 
\be
y \rightarrow +\infty \; : \; \varphi(y) \sim a \; y^{1-\omega}
\hspace{0.3cm} \mbox{with} \; 0 \leq \omega \leq 1 \hspace{0.2cm} ,
\hspace{0.2cm} a>0 
\label{phiom}
\ee
and to display a singularity at a small negative value of $y$,
\be
y \rightarrow y_s^+ \; : \; \varphi(y) = - a_s \; \Gamma(\omega_s)
\; (y-y_s)^{-\omega_s}
\label{ys}
\ee
where we neglected less singular terms (note that this
behavior has indeed been observed in the quasilinear regime,
Bernardeau 1992). Taking advantage of these assumptions, one obtains 
(Balian \& Schaeffer 1989),
\ba
x \ll 1 \,: && {\displaystyle  h(x) \sim
\frac{a(1-\omega)}{\Gamma(\omega)} \; x^{\omega-2} } \\  
x \gg 1 \,: && {\displaystyle h(x) \sim a_s \; x^{\omega_s-1} \; e^{-x/x_s} } 
\label{has} 
\ea
with $x_s=1/|y_s|$. Hence, using (\ref{Phx}) we see that the density
probability distribution $P(\delta_R)$ shows a power-law behavior
from $(1+\delta_R) \sim \xia^{\;-\om/(1-\om)}$ up to $(1+\delta_R)
\sim x_s \xia$ with an exponential cutoff above $x_s \xia$. It implies
in particular that the function $h(x)$ measured
in numerical simulations can give rise to constraints on the
cumulant generating function from the inverse relation,
\be
\varphi(y) = \int_0^{\infty} \; \left( 1 - e^{-xy} \right) \; h(x) \;\d x
\label{phih}
\ee
Note that $h(x)$
depends on the power-spectrum and, in the absence of a reliable
theory for describing the nonlinear regime, it has to be obtained from
numerical simulations. 
%In our case, we choose the scaling function measured for
%$n=-2$ by Valageas et al. (2000) (where $n$ is the slope of the linear
%power-spectrum) because on the angular scales we consider ($0.1' \la
%\theta \la 2'$) most of the contributions to the weak lensing effects
%come from scales where $n \simeq -2$ (see also Valageas 2000b). 
This  is  the  case  in  particular  for  the  evolution  of  the  two-point
correlation function,  or equivalently of the power-spectrum.  To this order
we use the  analytic formulae obtained by Peacock \&  Dodds (1996) from fits
to N-body simulations.

Note   that   the  relation   (\ref{hphi})   holds   independently  of   the
stable-clustering  ansatz.  However,  if  the  latter is  not  realized  the
generating function $\varphi(y)$ depends on  time (and scale). Then, most of
the results we obtain in the next  sections still hold but one needs to take
into  account the  evolution with  redshift of  $\varphi(y)$. That  would be
necessary in  particular if  one wants to  describe the transition  from the
quasilinear regime  to the strongly  nonlinear regime.  In the  following we
assume that the  stable-clustering ansatz is valid, so  that $\varphi(y)$ is
time-independent. As mentioned above this is consistent with the results of
numerical simulations.

\subsection{Minimal tree-model}
\label{Minimal tree-model}

If one is interested in the statistics of the top-hat filtered
convergence, it is reasonable to assume that (Valageas 2000a,b),
\be
\varphic(y)\approx \varphi_{3D}(y).\label{MfApprox}
\ee
In case of the aperture mass statistics however the filtering 
scheme is too intricate (with both positive and negative weights) 
to make such an assumption, and in particular
the resulting values for $S_p$ depend crucially on the
geometrical dependences of the $p$-point correlation functions. 
We are thus forced to adopt a specific model for the correlation
functions, and the one we adopt is obviously 
consistent with the stable-clustering ansatz.  

A popular model for the $p-$point correlation functions in the
non-linear regime is to consider a ``tree-model'' (Schaeffer 1984,
Groth \& Peebles 1977) where $\xi_p$ is expressed in
terms of products of $\xi_2$ as: 
\be 
\xi_p({\bf r}_1, ... ,{\bf r}_p) = \sum_{(\alpha)} Q_p^{(\alpha)}
\sum_{t_{\alpha}} \prod_{p-1} \xi_2({\bf r}_i , {\bf r}_j) 
\label{tree}
\ee
where $(\alpha)$ is a particular tree-topology connecting the $p$
points without making any loop, $Q_p^{(\alpha)}$ is a parameter
associated with the order of the correlations and the topology
involved, $t_{\alpha}$ is a particular labeling of the topology
$(\alpha)$ and the product is made over the $(p-1)$ links between the
$p$ points with two-body correlation functions. A peculiar case of the
models described by (\ref{tree}) is the ``minimal tree-model''
(Bernardeau \& Schaeffer 1992, 1999, Munshi, Coles \& Melott 1999)
where the weights $Q_p^{(\alpha)}$ are given by: 
\be
Q_p^{(\alpha)} = \prod_{\mbox{vertices of } (\alpha)} \nu_q
\label{mintree}
\ee
where $\nu_q$ is a constant weight associated to a vertex of the tree
topology with $q$ outgoing lines. Then, one can derive the generating
function $\varphi(y)$, defined in (\ref{phidef}), or
the coefficients $S_p$, from the parameters $\nu_p$ introduced in
(\ref{mintree}) which completely specify the behavior of the
$p-$point correlation functions.

In this case the cumulant generating function is given for 3D filtering by,
\be
\chi(\lambda) = \sum_{p=1}^{\infty} {\lambda^p\over p!} 
\int \d^3\vr_1 .. \d^3 \vr_p \; w(\vr_1) .. w(\vr_p) \; \xi_p(\vr_1,..,\vr_p)
\label{chiNL3Ddef}
\ee
where $w$ corresponds to the filter choice.
In the case of the minimal tree-model, where the
$p-$point correlation functions are defined by the coefficients
$\nu_q$ from (\ref{tree}) and (\ref{mintree}), it is possible to
obtain a simple implicit expression for the function $\chi(\lambda)$
(see Bernardeau \& Schaeffer 1992; Jannink \& des Cloiseaux 1987): 
\ba
{\displaystyle \chi(\lambda) } & = & {\displaystyle
\lambda \int {\d^3\vr} \; w(\vr) \; \left[ \zeta[ \tau(\vr)] - \frac{\tau(\vr)
\zeta'[\tau(\vr)]}{2} \right] } \label{chiNL3D}\\ 
{\displaystyle \tau({\bf r}) } &=&
{\displaystyle \lambda \int {\d^3\vr'} \; w(\vr') \; \xi_2(\vr,\vr') \;
\zeta'[\tau(\vr')] } 
\label{tauNL3D}
\ea
where the function $\zeta(\tau)$ is defined as the generating
function for the coefficient $\nu_p$,
\be
\zeta (\tau) = \sum_{p=0}^{\infty} \frac{(-1)^p}{p!} \; \nu_p \;
\tau^p \hspace{0.4cm} \mbox{with} \hspace{0.4cm} \nu_0 = \nu_1 = 1.  
\label{zetaNL}
\ee
The function $\chi(\lambda)$ obviously depends on the choice of filter
through the function $w$. For a top-hat filter, it would simply be
a characteristic function normalized in such a way that
\be
\int\d^3\vr \; w(r)=1.
\ee

A simple ``mean field'' approximation which provides very good results
in case of top-hat filter
(Bernardeau \& Schaeffer 1992) is to integrate $\tau({\bf r})$ over
the volume $V$ in the second line of the system (\ref{chiNL3D}) and then
to approximate $\tau({\bf r})$ by a constant $\tau$. This leads to the
simple system: 
\ba
\varphi_{3D}(y)&=&y\left[ \zeta(\tau)-\frac{\tau\;\zeta'(\tau)}{2}\right] 
\label{phiNL3DMF} \\
\tau&=& - y\;\zeta'(\tau)
\label{tauNL3DMF}
\ea
Then, the singularity of $\varphi(y)$, see (\ref{ys}), corresponds to
the point where the $|\d y/\d\tau|$ vanishes. Note that
$\zeta(\tau)$ is regular at this point and that the singularity is
simply brought about by the form of the implicit system
(\ref{tauNL3DMF}) as observed in Bernardeau \& Schaeffer
(1992). Making the approximation (\ref{phiNL3DMF}-\ref{tauNL3DMF})
for both $\varphi_{3D}(y)$ and $\varphic(y)$ leads to the
approximation (\ref{MfApprox}) which is thus natural for the minimal
tree model. 

In the case of a compensated filter such a simple mean field
approximation however cannot
be done. It is in particular due to the fact that the weights given to $\tau$ 
then strongly depend on the radius distance. Before we go to this point we need
first to take into account the projection effects.

\subsection{Projection effects for the minimal tree-model}
\label{Case of the minimal tree-model}

As noted in T\'oth et al. (1989) and analyzed in detail 
in Valageas (2000b), we know that the tree structure assumed for the 3D
correlation functions is preserved (except for one final integration along the
line-of-sight) for the projected density. Indeed, inserting (\ref{tree}) in
(\ref{cum1}) we obtain:
\ba
{\displaystyle \lag \deltap ({\vec \vartheta}_1) .. \deltap({\vec
\vartheta}_p) \rag_c = \sum_{(\alpha)} Q_p^{(\alpha)} \sum_{t_{\alpha}}
\int_0^{\drad_s} \d\drad_1 \; F^p } \nonumber\\ 
{\displaystyle \hspace{3cm} \times \;
\int_{-\infty}^{\infty} \prod_{i=2}^{p} \d\drad_i \; \prod_{p-1} \xi_2({\bf x}_a ,
{\bf x}_b) } 
\label{tree1}
\ea
where we noted ${\bf x}_a = (\drad_a , \De_1 {\vec \vartheta}_a)$. It
can be noted that in the small angle approximation the weight applied
to each diagram depends on their order $p$ only and not on their
geometrical decompositions.
As a consequence the projected $p-$point correlation function can be written,
\ba
\lag \deltap ({\vec \vartheta}_1) \dots \deltap({\vec \vartheta}_p)
\rag_c &=&\nonumber\\ 
&&\hspace{-2cm}\int_0^{\drad_s} \d\drad \; F^p \; \om_p({\vec
\vartheta}_1,\dots,{\vec \vartheta}_p;z),
\label{deltapomp}
\ea
where the two-dimensional $p-$point functions $\om_p({\vec
\vartheta}_1,...,{\vec \vartheta}_p;z)$ have the same tree-structure
as the three-dimensional $p-$point correlation functions $\xi_p$,
\be
\om_p({\vec \vartheta}_1,\dots,{\vec \vartheta}_p;z) = \sum_{(\alpha)}
Q_p^{(\alpha)} 
\sum_{t_{\alpha}} \prod_{p-1} \om_2({\vec \vartheta}_a,{\vec \vartheta}_b;z)
\label{treeom}
\ee
with:
\be
\om_2({\vec \vartheta}_1,{\vec \vartheta}_2;z) = \int\frac{\d^2
\vk}{(2\pi)^2} \; P(k,z)\; J_0 \left( k \De | {\vec \vartheta}_1 -
{\vec \vartheta}_2| \right) . 
\label{om2}
\ee
Here $J_0$ is the Bessel function of order 0. For convenience we also note
$\oma_2(z)$ the angular average of $\om_2({\vec \vartheta}_1,{\vec
\vartheta}_2;z)$,
\be
\oma_2(z)=\int{\d^2{\vec\vartheta}_1}{\d^2{\vec\vartheta}_2}\,U(\vec\vartheta_1)\,
U(\vec\vartheta_2)\,\om_2({\vec \vartheta}_1,{\vec \vartheta}_2;z),
\ee
which, expressed in terms of the power spectrum gives,
\be
\oma_2(z)=\int_0^{\infty}{\d^2\vk\over (2\pi)^2}P(k)\,W^2(k\De\theta).
\ee

Thus, we see that the correlation functions of the projected density $\deltap$
itself do not show an exact tree-structure as the underlying 3D correlation functions
$\xi_p$. Nevertheless, as seen in (\ref{deltapomp}) they are given by
one simple 
integration along the line-of-sight of the 2D $p-$point functions $\om_p$ which
exhibit the {\it same tree-structure} as their 3D counterparts $\xi_p$. This means
that we can still use the techniques developed to deal with such tree-models. In
particular, in the case of the minimal tree-model (\ref{mintree}) we will be able
to take advantage of the resummation (\ref{chiNL3D}-\ref{tauNL3D}).

Once again, it is interesting to note that for power-law spectra,
$P(k) \propto k^n$, the angular and the redshift
dependences of $\om_p$ can be factorized so that the
correlation functions of the projected density $\deltap$ itself now exhibit a
(new) tree-structure. Then, the 2-point function reads,
\ba
\lag \deltap ({\vec \vartheta}_1) \deltap({\vec
\vartheta}_2) \rag &=& |{\vec \vartheta}_1 - {\vec \vartheta}_2
|^{-(n+2)} \nonumber\\
&&\hspace{-3cm}
\times\int_0^{\infty} \frac{\d^2 \vl}{(2\pi)^2} \; l^{n} \; J_0(l) 
\int \d\drad \; F^2(\drad) \; \frac{P_0}{k_0^n} \; \De^{-(n+2)} 
\label{factn}
\ea
while the high-order $p-$point functions $\lag \deltap ({\vec \vartheta}_1) ..
\deltap({\vec \vartheta}_p) \rag_c$ follow the tree-structure (\ref{tree}) with
the projected weights $Q_{p,\,{\rm proj.}}^{(\alpha)}$:
\be
Q_{p,\,{\rm proj.}}^{(\alpha)}= Q_{p}^{(\alpha)} \; \frac{ \int \d\drad \, F^p \,
\left[ P_0(z) \, \De^{-(n+2)} \right]^{p-1} }{ \left[ \int \d\drad \, F^2
\, P_0(z) \, \De^{-(n+2)} \right]^{p-1}} .
\label{Qproj}
\ee
Note that the relation $Q_{p}^{(\alpha)} \leftrightarrow Q_{p,\,{\rm
proj.}}^{(\alpha)}$ depends on the slope $n$ of the power-spectrum. On the other
hand, if the initial tree-model for the 3D correlation functions is the minimal
tree-model (\ref{mintree}) we can see that the projected tree-structure
(\ref{Qproj}) is {\em not an exact minimal} tree-model\footnote{This
is due to the fact that the numerator in the r.h.s. of (\ref{Qproj})
cannot be factorized in the form $A \, B^{p-1}$. A simple way to check
that $Q_{p,\,{\rm proj.}}^{(\alpha)}$ cannot be written in terms of
new parameters $\nu_{q,\,{\rm proj.}}$ as in (\ref{mintree}) is to
consider the ``snake'' topology where $Q_{p}^{({\rm snake})} = \nu_1^2
\, \nu_2^{p-2}$.} which would be expressed in terms of a new
generating function $\zeta_{\rm proj.}(\tau)$.  In other words
$\varphip(y)$ cannot be built from a tree structure whereas
$\varphic(y)$ can, and with the {\em same} vertex generating function
$\zeta(\tau)$ defined in (\ref{zetaNL}).

As a consequence, in the nonlinear regime, the relation (\ref{phiproj}) is to
be used with,
\ba
\varphi_{\rm cyl.}(y) & = &
{\displaystyle y \int {\d^2\vartheta} \; U({\vec \vartheta}) \left[ \zeta[
\tau({\vec \vartheta})] - \frac{\tau({\vec \vartheta}) \zeta'[\tau({\vec
\vartheta})]}{2} \right] } \label{phicylNL2D} \\ 
\tau({\vec \vartheta})  &=&
{\displaystyle - y \int {\d^2 \vartheta'} \; U({\vec \vartheta}') \;
\frac{\om_2({\vec \vartheta} , {\vec \vartheta}';z)}{\oma_2(z)} \; \zeta'[\tau({\vec
\vartheta}')] }
\label{taucylNL2D}
\ea
where $\zeta(\tau)$ is the 3D vertex generating function.
Note that this function depends on $z$ through $\om_2({\vec \vartheta}
, {\vec \vartheta}';z)$ and $\oma_2(z)$.
Note also that in $y=0$ the expansion of the generating function
$\varphi_{\rm cyl.}(y)$ is  $\varphi_{\rm cyl.}(y) = -y^2/2 + \dots$.

\begin{figure}
{\epsfxsize=8cm {\epsfbox{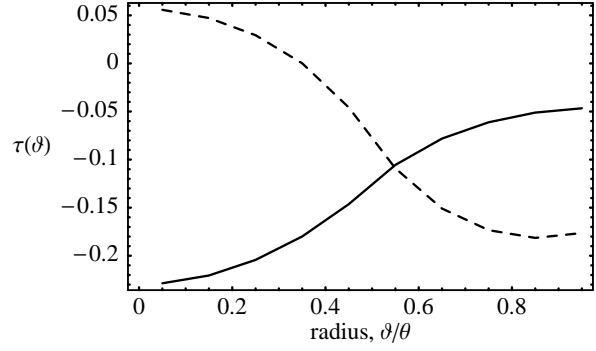}} }
\caption{Profile of the $\tau$ function as a function of the angular
radius. The two plots correspond to the two singularities, $y_{s-}$ for the
solid line and $y_{s+}$ for the dashed line.} 
\label{TauProfil}
\end{figure}

We have computed the resulting shape of the generating function
in such a model for various cases. For comparison with the previous quasi-linear
case we assume here the power spectrum to follow a power law behavior with index
$n=-1.5$. The vertex generating function is assumed to be given by
\be
\zeta(\tau)=(1+\tau/\ka)^{-\ka}\label{zetaNLform}
\ee
with $\ka\approx 0.5$ (in the parameterization of $\zeta$ we followed
the traditional notation and used $\kappa$
as a simple free parameter. It is not to be confused
with the local convergence).
In Fig. \ref{TauProfil} we present typical profiles obtained for $\tau(\vartheta)$.
We see that it is regular in $\vartheta$. In particular it does not exhibit
discontinuities nor abnormal behavior near the singular values
of $y$. We also found that the results we obtain are very
robust regarding to the number of shells used to describe the integral in
$\vartheta$: with 2 cells only the description of $\varphic$
is already very accurate.

The choice of the value of $\kappa$ in (\ref{zetaNLform}) relies a
priori on numerical results. The EPT (Colombi et al. 1997) or more
convincingly the HEPT (Scoccimarro \& Frieman 1999) provide however a
convenient frame which can be used to predict the value of $\kappa$.
This can be done for instance by identifying the predicted values for
the skewness both from the form (\ref{zetaNLform}) and HEPT. 
Indeed in our model we have,
\be
S_3(\kappa)=3{(1+\kappa)\over \kappa}
\ee
whereas, in HEPT, $S_3$ is related to the initial power spectrum index,
\be
S_3^{\rm HEPT}(n)=3{4-2^n\over 1+2^{n+1}},
\ee
which leads to,
\be
\kappa\approx {\frac{\left( 1 + {2^{1 + n}} \right) }
 {3\,\left( 1 - {2^n} \right) }}.
\label{kappan}
\ee
In the numerical applications presented in the following we will use
this scheme. In particular, $n=-1.5$ leads to $\kappa=0.88$. On the other hand, at the angular scale $\theta =4'$ which we consider below the local slope of the linear power-spectrum is $n \simeq -2.2$ which leads to $\kappa \simeq 0.6$.

The properties of $\varphic(y)$ we get in this regime are very similar to
those obtained for the quasilinear regime. In particular we found that the
function $\varphic(y)$ exhibits 2 singular points on the real axis. As
for the quasilinear regime this behavior is due to the implicit equation
in $\tau$ and not to peculiar choice of the vertex generating function.
In Table \ref{SingTab} we summarize the parameters that describe the
singularities of $\varphic(y)$ in different regimes. It appears, as expected,
that the singularities are closer to the origin. This is to be expected
since the nonlinearities contained in $\varphic(y)$ are stronger in the
nonlinear regime compared to the quasilinear regime. 

\begin{table*}
\begin{center}
\caption
{Values of $S^{\rm cyl.}_3$ and singularity positions of $\varphic(y)$
for $U_{BV}$ filter.}
\label{SingTab}
\begin{tabular}{rlllll}
  & $S^{\rm cyl.}_3$ & $y_{s-}$ & $y_{s+}$ & $a_{s-}$ & $a_{s+}$ \\\hline 
quasi-linear, $n=-1.5$           & 2.53 & -0.092 & 0.159 & 1.20 & 0.74 \\
non-linear, $\ka=1.0 $, $n=-1.5$ & 4.19 & -0.062 & 0.130 & 0.85 & 0.41 \\
non-linear, $\ka=0.88$, $n=-1.5$ & 4.49 & -0.057 & 0.121 & 0.83 & 0.40 \\
non-linear, $\ka=0.50$, $n=-1.5$ & 6.32 & -0.040 & 0.084 & 0.65 & 0.31
\end{tabular}
\end{center}
\end{table*}

Similarly to the quasilinear regime it is also possible
to define an effective vertex generating function from which 
$\varphic(y)$ can be built and which reproduces its 
singular points,
\ba
\zeta_{\rm eff.}(\tau)&=&-\tau + 1.4966\,\tau^2 - 11.6982\,\tau^3 +
21.528\,\tau^4 - \nonumber\\
&&\ \ \ \ \ \ \ \ \ 77.1899\,\tau^5.
\ea
The result given here has been obtained for $\kappa=0.88$ in Eq. (\ref{zetaNLform}).

\section{Statistics of the aperture mass $\Map$}
\begin{figure}
\epsfxsize=8cm {\epsfbox{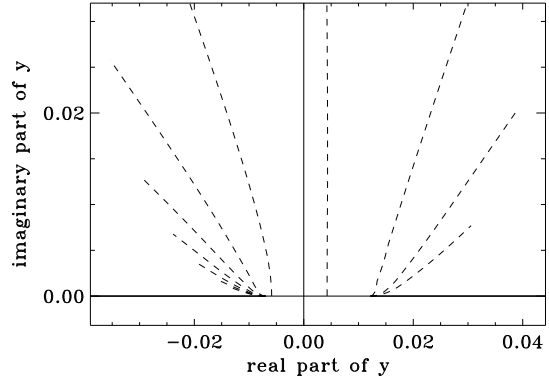}} 
\caption{Integration paths in the $y$ complex plane. The paths (dashed
lines) are dynamically built so that the argument of the exponential in
Eq. (\ref{PMapFinal}) is always a negative real number. The thick half
straight lines represent the locations of the non-analytic parts of $\varphip(y)$ in this plane. Of course, the paths must not cross these branch cuts.}
\label{Ypath}
\end{figure}

\label{Statistics of the aperture mass}

It now suffices to plug the numerical expressions
we have obtained for $\varphic(y)$ in (\ref{phiproj}) to get 
the shape of the $\Map$ PDF.
More precisely we have,
\ba
P(\Map)\d\Map&=&\vert\kappa_{\rm min}\vert\,\d\Map
\int{\d y\over 2\pi\ii\sigma^2}\nonumber\\
&&\hspace{-1.6cm}
\times\exp\left[-{\vert\kappa_{\rm min}\vert^2
\varphip(y)\over \sigma^2}+{y\vert\kappa_{\rm min}\vert\Map\over \sigma^2}\right].
\label{PMapFinal}
\ea
where $\sigma$ is the variance of the aperture mass.

In (\ref{PMapFinal}) the integral over $y$ has to be made in the
complex plane. The integration path in the $y$ plane is built in such
a way that the argument of the exponential  is always a real negative
number thus avoiding oscillations (see Fig. \ref{Ypath}). Moreover, one must make sure that the integration path does not cross the branch cuts of $\varphic(y)$. The 
singularities of $\varphic(y)$ induce non-analytic parts for
$\varphip(y)$ as well. They are located at positions,
\be
y>y_{s+}^{\rm proj}\ \ {\rm and}\ \ y<y_{s-}^{\rm proj}
\ee
with
\be
y_{s\pm}^{\rm proj}=y_{s\pm}/\max[F(\drad)\psi(\drad)]
\ee
where the maximum value of $F(\drad)\psi(\drad)$ is taken along the
line-of-sight (and is indeed finite). As noticed in Valageas (2000a) the exponent $\omega_s$ of the singularity of $\varphic(y)$ (as defined in (\ref{ys})) leads to the exponent $\omega_s-1/2$ for the projected generating function $\varphip(y)$. However, in both the quasi-linear and highly non-linear regimes we have $\omega_s=-3/2$, see (\ref{omsL}). In this case, as shown in App. A, for $y \rightarrow y_{s\pm}^{\rm proj}$ the singularity is of the form $\varphip(y) \sim (y-y_{s\pm}^{\rm proj})^2 \ln |y-y_{s\pm}^{\rm proj}|$.

The existence of these branch cuts is directly responsible for two exponential cut-offs in
the shape of the PDF of $\Map$,
\be
P(\Map)\sim \exp\left({\vert\kappa_{\rm
min}\vert\Map\over \sigma^2}\,y_{s\pm}^{\rm proj}\right).
\label{tailPMap}
\ee
It can be noted that the $\Omega$ dependence of $\kappa_{\rm min}$
will induce a strong $\Omega$ dependence in the position of the
exponential cut-offs. The variation of $\kappa_{\rm min}$ with $\Omega$
is thus to be compared with the theoretical uncertainties on $y_{s\pm}$.

\subsection{The PDF shape}

\begin{figure}
\epsfxsize=8cm {\epsfbox{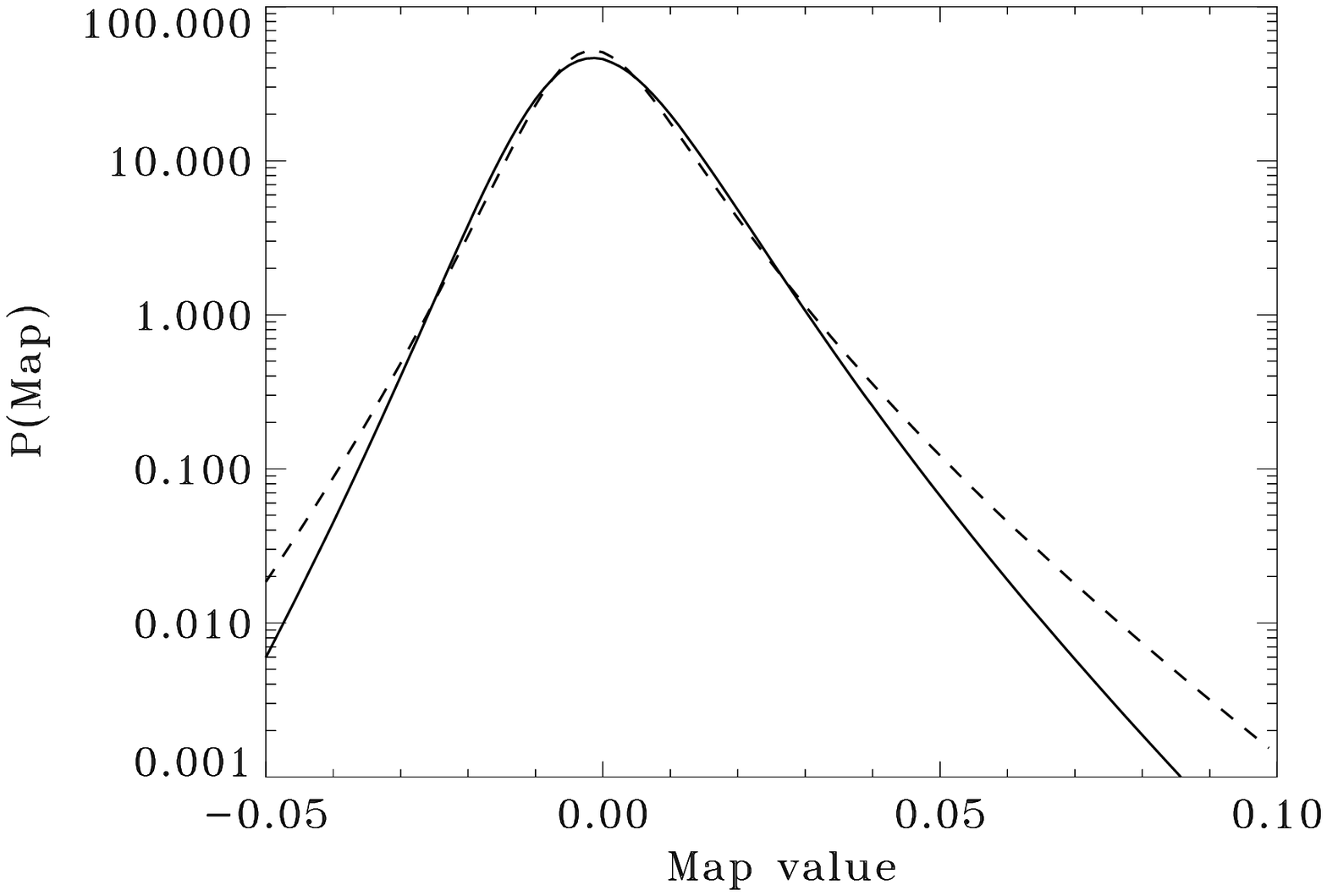}} 
\epsfxsize=8cm {\epsfbox{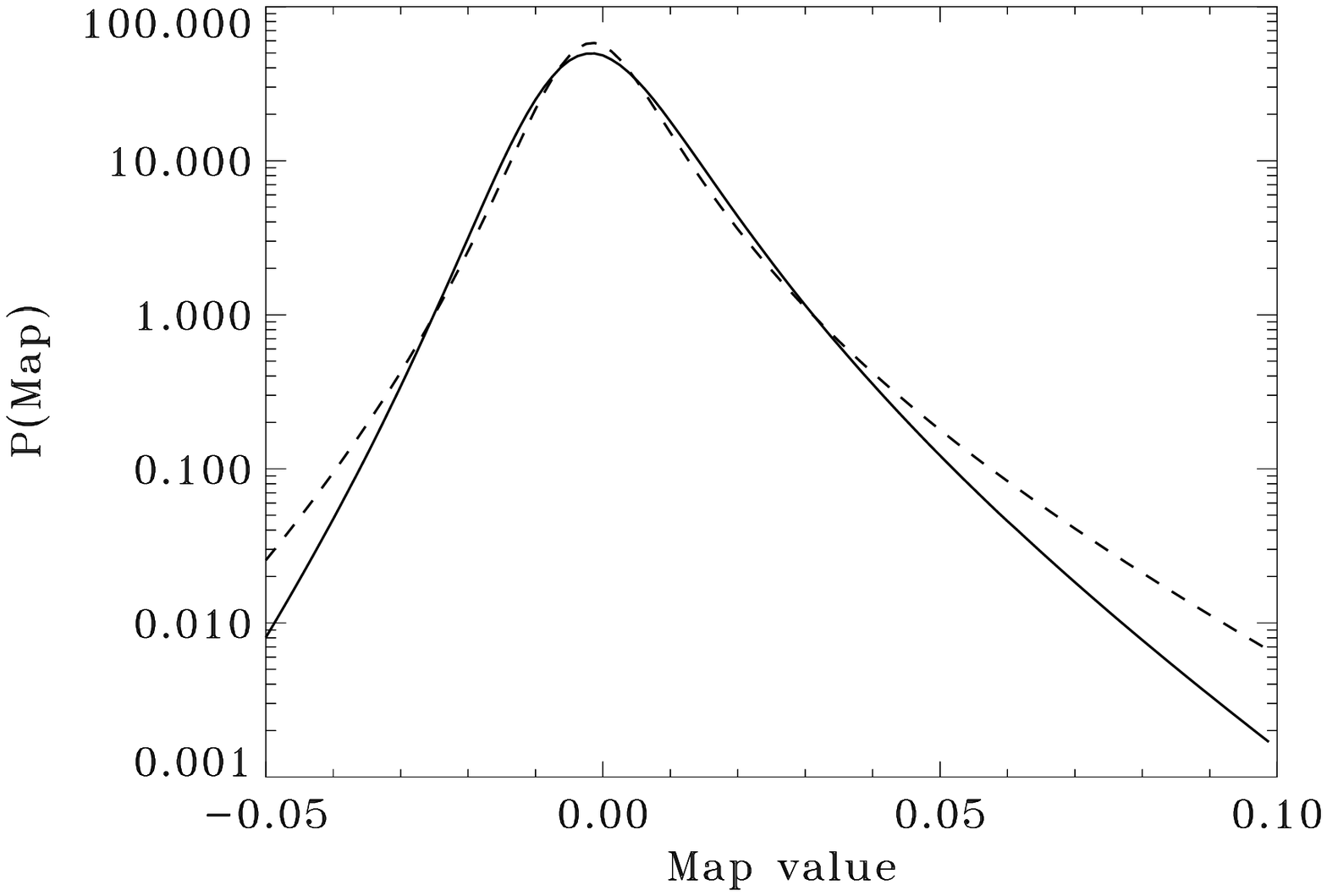}} 
\caption{Shape of the one-point PDF of the aperture mass from the
quasi-linear regime model (top panel) and in the Nonlinear regime (bottom
panel).  The sources are at redshift unity, the
variance of $\Map$ is $0.01$, for the filter given by Eq. \ref{Ubv}.
the solid lines correspond to Einstein-de Sitter case, the dashed lines
to flat universe with $\Om=0.3$.} 
\label{NumPDF}
\end{figure}

In Figs.  \ref{NumPDF} we present the  resulting shape of the  one point PDF
obtained   for  different   cases.  They   have  been   obtained   from  the
parameterization  of $\varphic(y)$  described in  the previous  sections. In
particular we assume a power  law spectrum with index $n=-1.5$. The variance
adopted for the  plots is $\sigma=0.01$. In these  investigations we did not
try to put a realistic source distribution, but we assume all the sources to
be  at  redshift unity.  However,  all  our results  can  be  extended in  a
straightforward fashion  for any redshift  distribution of the  sources (the
latter  is simply  absorbed  by  a redefinition  of  the selection  function
$F(\drad)$). Note also that we  present the PDF for the correctly normalized
aperture  mass $\Map$,  that is  without dividing  the local  convergence by
$\kappa_{\rm min}$. We can check in  Fig. \ref{NumPDF} that the tails of the
PDF are  stronger in the non-linear  regime than in  the quasi-linear regime
independently  of  the  variance  $\sigma$   (which  is  the  same  in  both
plots).  This  is  related  to  the  smaller  values  of  the  singularities
$|y_{s\pm}^{\rm proj}|$ in  the non-linear case, as shown  by the expression
(\ref{tailPMap}).  Of  course, this  is  due to  the  smaller  value of  the
singularity $|y_s|$ of  the 3D density field in the  non-linear regime (in a
similar fashion the coefficients $S_p$ are larger).

\begin{figure}
\epsfxsize=8cm {\epsfbox{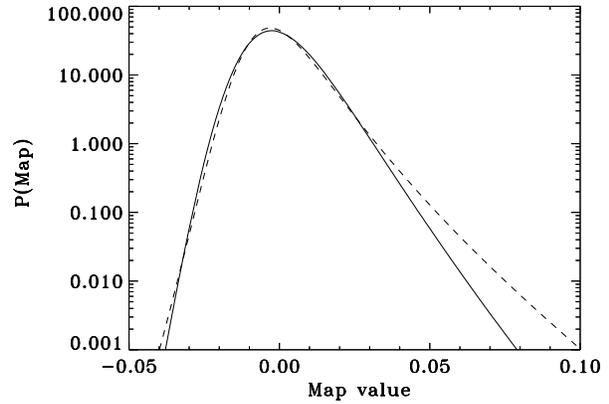}} 
\caption{Same as previously with the nonlinear model but with
cell radius ratio of 10 instead of 2.} 
\label{NumPDFp}
\end{figure}

In Fig. \ref{NumPDFp} we show that the positions of the cut-off depend
crucially on the window shape. In particular it is clear that when the
disc radius ratio is larger the PDF is more asymmetric and bears more resemblance
with the $\kappa$-PDF for a top-hat window function. Indeed, when the inner disk is much smaller than the outer radius the fluctuations of $\Map=\kappa_1 - \kappa_2$ are dominated by those of the convergence $\kappa_1$ which corresponds to this small inner window while $\kappa_2$ which is governed by larger scales shows lower amplitude fluctuations.

\subsection{The $\Om$ dependence of the PDF}

It has  been stressed in the  literature (e.g., Bernardeau et  al. 1997) that
the non-Gaussian properties of the  convergence maps are expected to exhibit
a strong  $\Om$ dependence. This is  due in particular  to the normalization
factor.  Such dependence  is  apparent in  $\kappa_{\rm  min}$ that  depends
crucially on  the value  of $\Om$.  This  property naturally extends  to the
shape of the  one-point PDF.  In particular it is important  to have in mind
that the $\Om$ dependence is  negligible in $\varphic(y)$ (for a fixed shape
of the power spectrum). The $\Om$ dependence is therefore entirely contained
in the projection  effect through the shape and  amplitude of the efficiency
function.

In  Fig.  \ref{NumPDF}  we  show   how  low  values  of  $\Om$  amplify  the
non-Gaussian features contained in the  PDF. Whether such a parameter can be
constrained  more  efficiently with  the  PDF  than  with simply  the  local
skewness is not  yet clear. Such a study is however  beyond the objective of
this paper and is left for further works.

\subsection{Comparison with numerical simulations}

\begin{figure*}
\begin{center}
\epsfxsize=8.1 cm \epsfysize=6 cm {\epsfbox{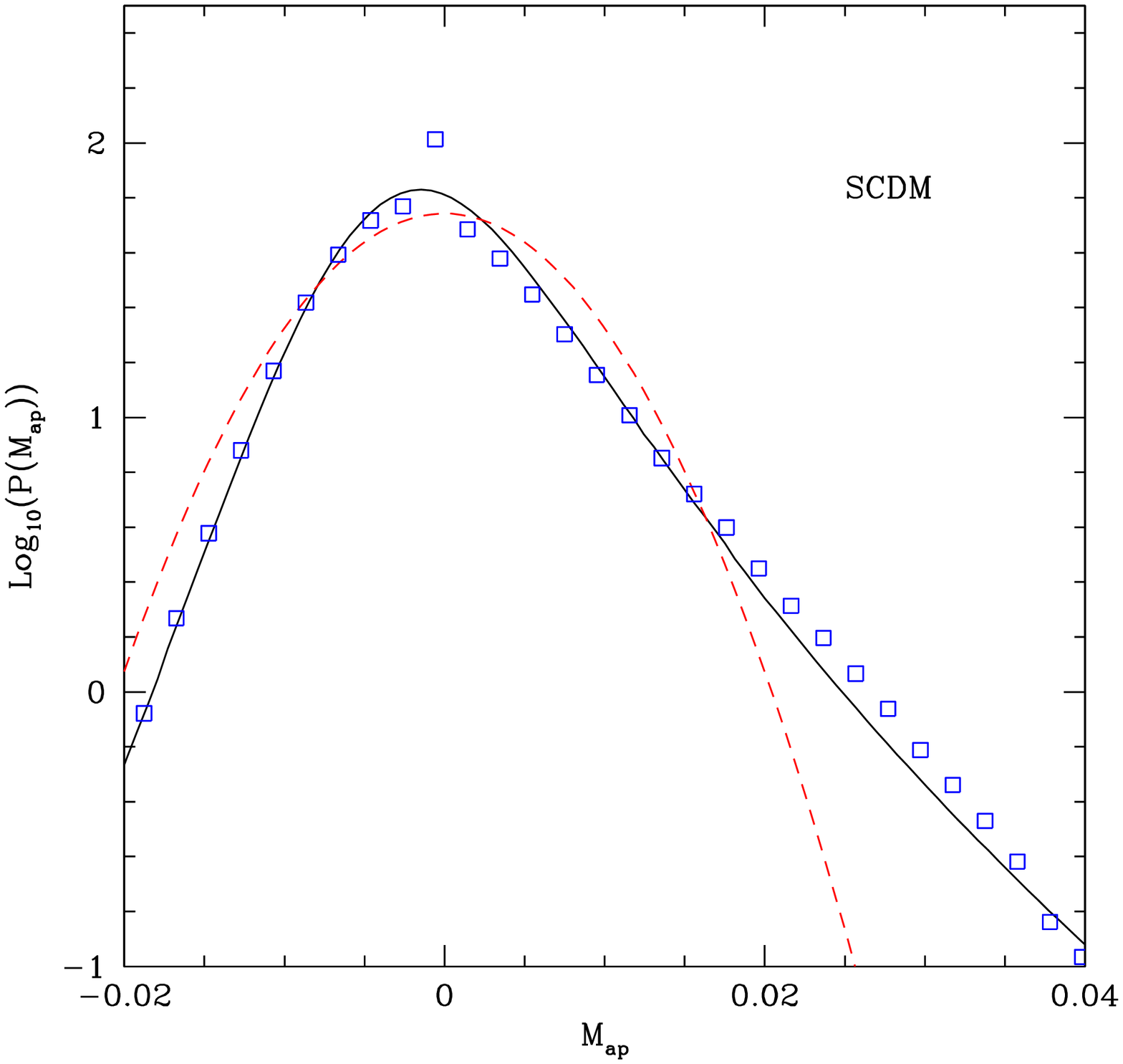}}
\epsfxsize=8.1 cm \epsfysize=6 cm {\epsfbox{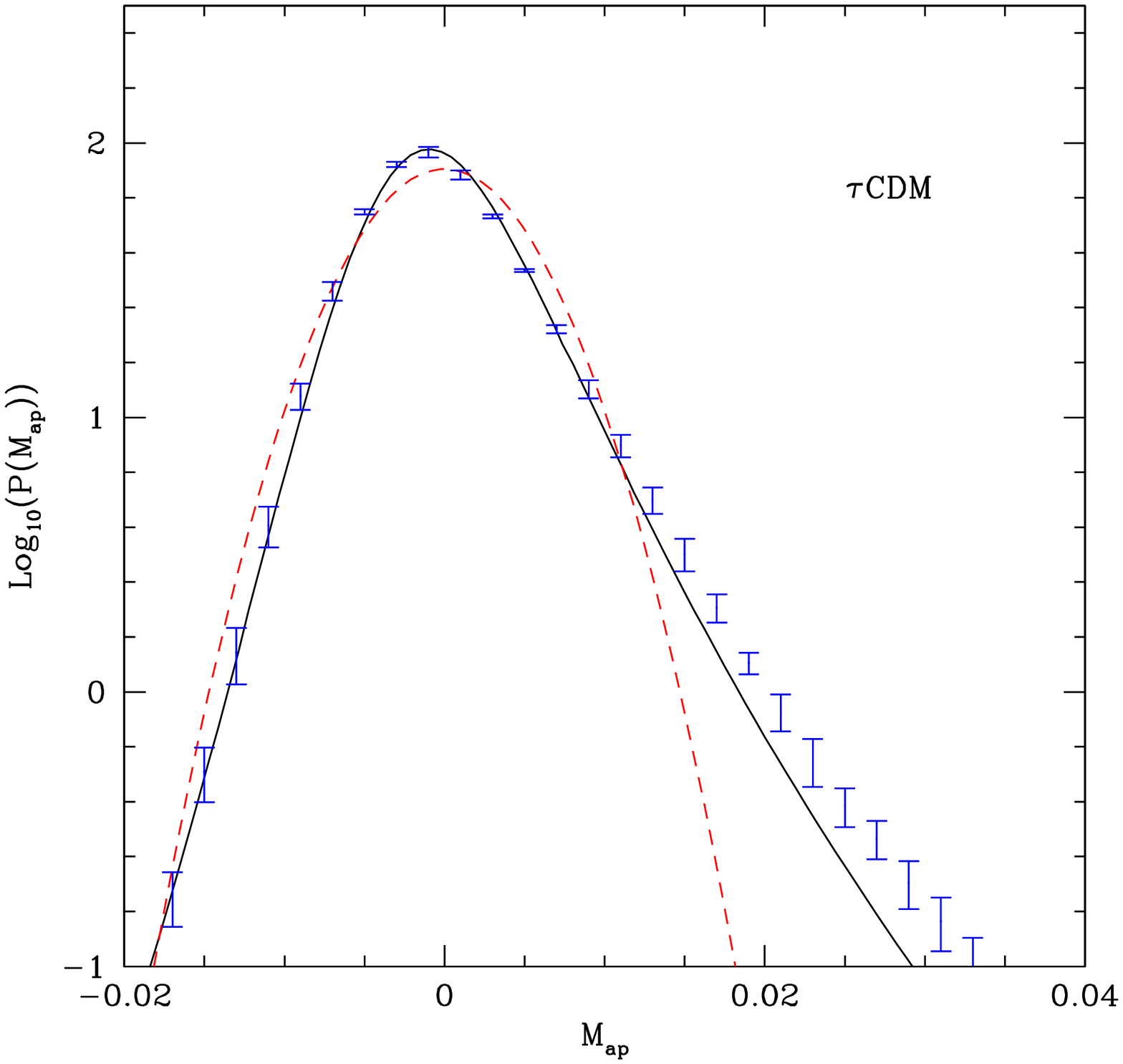}}\\
\epsfxsize=8.1 cm \epsfysize=6 cm {\epsfbox{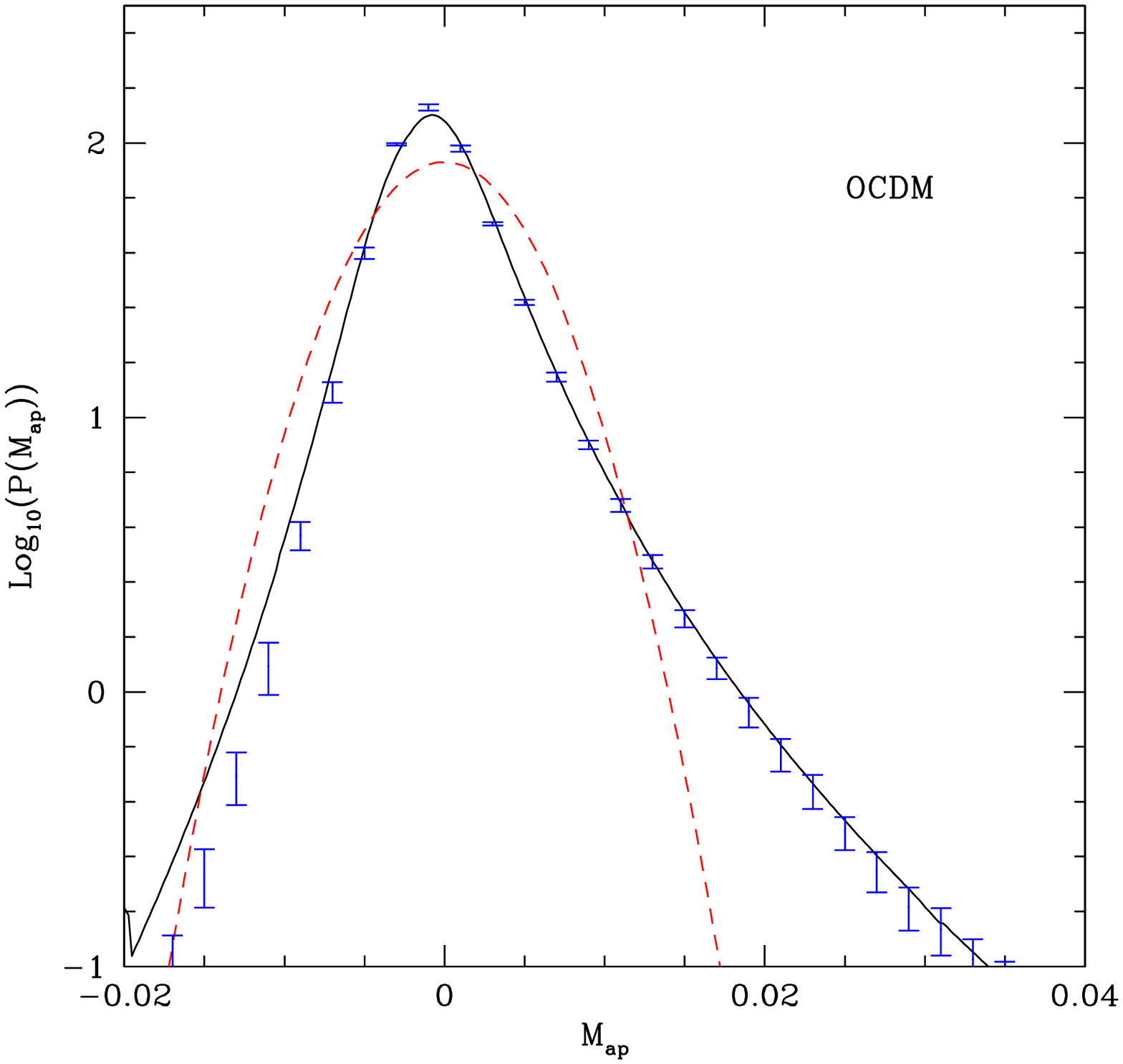}}
\epsfxsize=8.1 cm \epsfysize=6 cm {\epsfbox{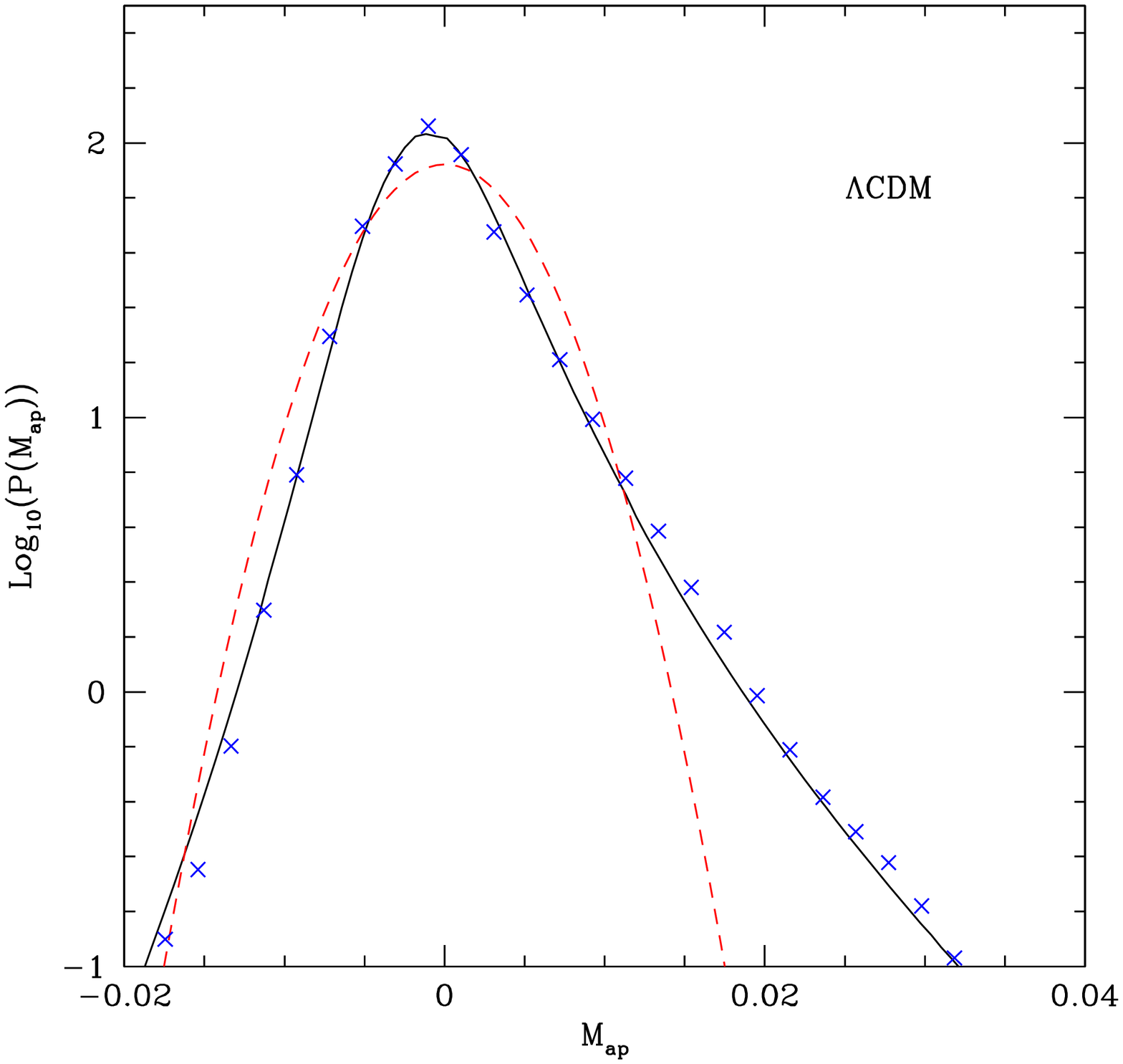}}
\end{center}
\caption{The  aperture mass  PDF  for a  source  at redshift  $z_s=1$ for  4
cosmologies and with  the angular window $\theta=4'$. The  dashed line shows
the Gaussian  which has the  same variance. The  points show the  results of
N-body simulations from  Jain, Seljak \& White (2000).  Results for sCDM and
$\Lambda$CDM models have been directly taken from Reblinsky et al. (1999).}
\label{figPlMap} 
\end{figure*}

Finally, we  compare our predictions for  the PDF $P(\Map)$  of the aperture
mass $\Map$  with the results of  N-body simulations (Jain,  Seljak \& White
2000) in Fig.  \ref{figPlMap}.  {\em Note  that for all these comparisons we
exclusively  use the  $u_S$ filter}.   We consider  the  cosmological models
defined in Tab.\ref{table1}: a standard  CDM (SCDM) and a $\tau$CDM scenario
in a  critical density  universe, a low-density  open universe (OCDM)  and a
low-density   flat   universe   with   a  non-zero   cosmological   constant
($\Lambda$CDM).   Here   $\Gam$  is  the   usual  shape  parameter   of  the
power-spectrum. We use the fit given by Bardeen et al. (1986) for $P(k)$. We
only consider the weak lensing distortions which affect a source at redshift
$z_s=1$, with angular window characteristic scale $\theta=4'$.

\begin{table*}
\begin{center}
\caption{Cosmological  models and  results obtained  with $U_S$  filter. The
sixth and seventh lines show the variance and the skewness of $P(\Map)$ from
numerical simulations  (Jain, Seljak  \& White 2000).  Results for  sCDM and
$\Lambda$CDM models  have been directly  taken from Reblinsky et  al. (1999)
and have been obtained with one  realization only. The eight and ninth lines
show the reconstructed PDF properties.}
\label{table1}
\begin{tabular}{|r|llll|}\hline
  & sCDM & $\tau$CDM (5 realizations) & OCDM (7 realizations)&  $\Lambda$CDM \\ \hline
$\Om$ & 1 & 1 & 0.3 & 0.3 \\
$\Ol$ & 0 & 0 &  0 &   0.7  \\
$H_0$ [km/s/Mpc] & 50 & 50 & 70 & 70 \\
$\sigma_8$ & 0.6 & 0.6 & 0.85 & 0.9 \\
$\Gam$ & 0.5 & 0.21 & 0.21 & 0.21 \\ \hline
$\sigma_{\Map}$, $\theta=4'$ & 0.00730 & $0.00542 \pm  0.00017$ & $0.00449 \pm 0.00017$ & 0.00495 \\ 
$s_3^{\Map}$, $\theta=4'$ & 159 & $ 258. \pm 38.  $        & $ 545. \pm 30.$
& 347 \\ \hline
$\sigma_{\Map}$, $\theta=4'$ & 0.00721 & $0.00497$ & $0.00469$ & 0.00477 \\ 
$s_3^{\Map}$, $\theta=4'$ & 148 & $ 180 $ & $ 395 $ & 300 \\ \hline
\end{tabular}
\end{center}
\end{table*}

%\begin{figure}
%\epsfxsize=8 cm {\epsfbox{NumPDF_Bhuv_EdS.ps}}
%\epsfxsize=8 cm {\epsfbox{NumPDF_Bhuv_Open.ps}}
%\epsfxsize=8 cm {\epsfbox{NumPDF_Bhuv_Diff.ps}}
%\caption{The  aperture mass PDF  for a  source at  redshift $z_s=1$  for two
%cosmologies (solid line: Einstein-de  Sitter case, dashed line: open Universe
%with $\Omega=0.3$) and with the  angular window $\theta=2'$. The points show
%the results of N-body simulations from Jain, Seljak \& White 2000}
%\label{figPlMap} 
%\end{figure}

In the  numerical calculations,  we discretize the  integral (\ref{phiproj})
over   redshift  and   we   solve  for   the   system  (\ref{phicylNL2D}   -
\ref{taucylNL2D}). That is  we take into account the  redshift dependence of
the  generating  function  $\varphi_{\rm  cyl.}(y)$. Moreover,  we  use  the
relation (\ref{kappan})  to get the  value of the parameter  $\kappa$, where
for  $n$  we take  the  local  slope of  the  linear-power  spectrum at  the
wavenumber $k=2/(\De(z_s/2)\theta)$.  This corresponds to  the Fourier modes
which  are   probed  by   the  filter  of   angular  radius   $\theta$,  see
Fig.\ref{FilterShape}. For $\theta=4'$ we obtain $n \simeq -2.2$ and $\kappa
\simeq 0.6$.

First, we can  check in Fig. \ref{figPlMap} that we  recover the right trend
for $P(\Map)$, with two asymmetric  tails for large $|\Map|$. In particular,
the exponential cutoff is stronger  for negative values of the aperture mass
than for  positive values.  We can  also note that the  PDF is significantly
different  from a  Gaussian as  it shows  a clear  exponential  cutoff, much
smoother than  the Gaussian falloff,  especially for large  positive $\Map$.
On the  other hand,  we note that  Reblinsky et  al. (1999) obtained  a good
match to  the tail of the PDF  $P(\Map)$ using a description  of the density
field  as  a collection  of  virialized  halos  (Kruse \&  Schneider  1999).
However,  such a method  is restricted  to the  far tail  of the  PDF (large
positive $\Map$)  while our approach provides  in principle a  model for the
full PDF $P(\Map)$. There seems to be a small discrepancy with the simulations for the $\tau$CDM scenario. It is not clear whether this is due to a limitation of HEPT or of our formulation. To clarify this problem one should test the statistics of the 3D density field and of the projected density in the same simulation. However, this is beyond the scope of our paper. Nevertheless, the overall agreement appears to be quite reasonable. Note that the shape of the PDF is governed by only one parameter $\kappa$, which is uniquely related to the local slope of the power-spectrum, independently of scale and of the cosmology. On the other hand, the inaccuracy of the numerical simulations in the tail of the PDF might be somewhat underestimated.

\begin{figure}
\epsfxsize=8.1 cm \epsfysize=6 cm {\epsfbox{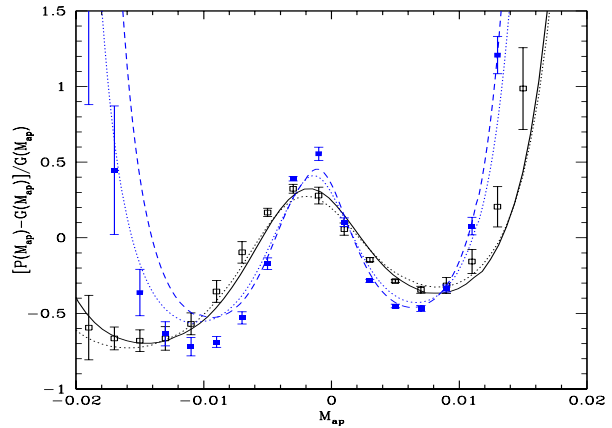}}
\caption{
The relative difference $[P(\Map)-G(\Map)]/G(\Map)$ of the aperture
mass PDF $P(\Map)$  with respect to the Gaussian  $G(\Map)$. The variance of
the Gaussian  is taken  from the simulations.  As in Fig.  \ref{figPlMap} we
consider   a  source   at   redshift  $z_s=1$   with   the  angular   window
$\theta=4'$. We display our results  for the $\tau$CDM (solid line) and OCDM
(dashed  line)  cosmologies.  The   dotted  lines  correspond  to  the  same
cosmologies with $\kappa=0.88$ (see main  text). The points show the results
of N-body simulations from Jain, Seljak \& White (2000).
}
\label{figPGausMap} 
\end{figure}

In  order to  see  more clearly  the  difference of  the  aperture mass  PDF
$P(\Map)$ with respect to the  Gaussian we display in Fig. \ref{figPGausMap}
the relative difference $P(\Map)/G(\Map)-1$.  Here $G(\Map)$ is the Gaussian
with the  same variance as the  numerical simulations. We can  check that we
recover   a  reasonable   agreement  with   the  numerical   results,  since
Fig. \ref{figPGausMap}  is directly related to Fig.   \ref{figPlMap}. To get
an  estimate  of  the sensivity  of  our  predictions  with respect  to  the
parameterization  (\ref{zetaNLform})  and  (\ref{kappan})  we also  plot  in
Fig.  \ref{figPGausMap} our  results for  the same  cosmologies when  we use
$\kappa=0.88$  (this would  correspond to  an initial  power  spectrum index
$n=-1.5$) in (\ref{kappan}).  We can see in the  figure that our predictions
are not too sensitive to $\kappa$ (for reasonable values of $\kappa$) within
the range $-0.015 < \Map < 0.015$ (the difference would look even smaller in
Fig. \ref{figPlMap}).  
In particular, the  variation with $\kappa$  of
our   results   is  much   smaller   than   the   difference  between   both
cosmologies. This suggests that one could  use the deviation of the PDF with
respect to a Gaussian to  estimate the cosmological parameters. A well-known
tool to  measure this signature is  the skewness but one  could devise other
statistics which  would take advantage of  the expected shape of  the PDF to
maximize their dependence on cosmology.  However, such a study is beyond the
scope of this article. We  can see in Tab.\ref{table1} that we underestimate
somewhat the skewness of $P(\Map)$. However, it is not clear whether this is
due  to the parameterization  (\ref{zetaNLform}) or  to the  use of  HEPT in
(\ref{kappan}).   We note  that the  tail  of $P(\Map)$  for large  negative
values of $\Map$ appears to be  slightly more sensitive to $\kappa$ than the
tail  at positive  $\Map$.   This could  be  related to  the  fact that  the
behaviour of $P(\Map)$ for $\Map  < -\sigma_{\Map}$ is more sensitive to the
detailed  properties  of  the  $p-$point  correlation  functions  (see
Valageas 2000c for a study of this point).

Finally, we note that  although $\theta=4'$ corresponds to non-linear scales
it is not very far from  the quasi-linear regime. However, the aperture mass
$\Map$ probes the  non-linear density field for filters  with larger angular
scale  than  the  convergence  $\kappa$.  Indeed, since  the  aperture  mass
involves compensated filters the contribution from low-$k$ modes is strongly
suppressed (see  Fig.  \ref{FilterShape}) so  that $P(\Map)$ is  governed by
the  properties of  the density  field at  the comoving  wavenumber  $k \sim
2/(\De  \theta)$.   In  contrast,  the  convergence $\kappa$  shows  a  more
important contribution  from larger wavelengths which implies  that in order
to  probe non-linear  scales  only one  must  set the  filter size  $\theta$
farther away into  the small-scale non-linear regime.  This  also means that
in principle  the aperture  mass could  be a more  convenient tool  than the
convergence  since  it should  be  easier  to  separate the  non-linear  and
quasi-linear regimes,  while for  an important range  of angular  scales the
convergence  should  be sensitive  to  the  transitory  regime between  both
domains. However, a  possible caveat is that the  statistics of the aperture
mass depend on the detailed behaviour of the $p-$point correlation functions
(and not on their average over spherical cells only), and therefore requires
a better understanding of them.

As discussed above, this property of the aperture mass to probe a narrow range of wavenumbers makes it easier to avoid the intermediate regime as one can select observation windows which are either in the quasi-linear or in the strongly non-linear regime. However, it would clearly be interesting to obtain a model which would also cover this transitory range. Unfortunately, this is rather difficult as one cannot use the simplifications which appear in the two extreme regimes. An alternative to a rigorous calculation would be to use an ad-hoc parameterization which would smoothly join the quasi-linear regime to the highly non-linear regime, for instance in the spirit of HEPT as described in van Waerbeke et al. (2000b) for the skewness of the convergence (note that our model for non-linear scales is based on a simple ansatz which is not rigorously derived). However, although the quasi-linear and strongly non-linear regimes share the same gross features, like the relation (\ref{phiproj}) which describes the projection effects, their detailed properties are different. In particular, although in both cases we have the scalings (\ref{scalings}) the correlation functions obtained in the quasi-linear regime are not given by a tree-model as in (\ref{tree}). This leads to the difference between the two-variable system 
(\ref{phicylql1})-(\ref{phicylql2}) and the integral relation (\ref{phicylNL2D})-(\ref{taucylNL2D}). Nevertheless, a simple prescription would be to recast the relations (\ref{phicylNL2D})-(\ref{taucylNL2D}) into the form (\ref{phicylql1})-(\ref{phicylql2}) by approximating the integrals over $\vartheta$ by the difference between two mean values, corresponding to the inner and outer regions of the filter $U_{BV}$ and characterized by two averages $\tau_1$ and $\tau_2$. Then, the shift from the quasi-linear to the highly non-linear regime would simply be described by a smooth interpolation of the generating function $\zeta(\tau)$, i.e. of the sole parameter $\kappa$. However, such a study is left for a future work as in this article we prefered to lay out the formalism needed to study the statistics of the aperture mass and to focus on the two regimes which have already been tested in details against numerical simulations, so as not to introduce a new specific parameterization.

\section{Conclusion}

In this article  we have described methods that  allow exact reconstructions
of the one-point PDF of the  local aperture mass in weak lensing maps. These
methods  do  take  into account  the  projection  effects  but not  all  the
nonlinear  couplings  between  the  local  density field  and  the  observed
distortions field such as lens-lens  coupling effects, or departure from the
Born approximation.

In  the course  of this  paper we  have examined  both the  quasi-linear and
non-linear regimes.  In particular, although the details of the calculations
are specific to  each case we have pointed out  the generic properties which
are common to both regimes and  the features brought about by the projection
effects.  For instance, in both  quasi-linear and non-linear domains the PDF
$P(\Map)$  should show two  asymmetric exponential  tails.  Our  methods are
quite  general and can  be extended  in a  straightforward fashion  to other
statistics.  In the  quasi-linear regime our approach can  be applied to any
filter which  is axisymmetric  while in the  non-linear regime there  are no
restrictions. In particular, in this latter case our results can be extended
to multivariate statistics (which can be obtained from filters which consist
of several disconnected parts).

We have  briefly investigated the dependence  of the PDF  $P(\Map)$ with the
shape of  the filter. Thus,  we have checked  that for filters with  a large
compensation radius we recover approximatly the shape of the PDF $P(\kappa)$
which is relevant for the top-hat filtered convergence.

Finally, we have checked that our predictions agree reasonably well with the
results of  available numerical simulations  (although we have  not included
any noise  effect at  this level) at  scale about  4' where the  data should
provide the  largest signal to noise  ratio (e.g. Jain \&  Seljak 1997).  In
particular,  we recover  the asymmetric  shape  of the  PDF.  Moreover,  our
approach provides  a prediction  for the full  shape of the  $P(\Map)$ while
earlier models  were restricted to  the positive tail  of the PDF.   We have
also shown that the difference between the PDFs obtained for two cosmologies
($\tau$CDM and OCDM)  is larger than the inaccuracy  of our predictions (due
to  parameterization we  need to  introduce  to describe  the underlying  3D
density field).   This suggests that our  results could be  used to estimate
the cosmological  parameters.  Thus,  in addition to  the skewness  which is
traditionally used to this purpose  one could take advantage of the expected
shape  of  the  PDF to  build  other  statistics  which would  maximize  the
dependence on the seeked parameters. Such a study is left for further work.

%On the  other hand, if the  cosmological parameters are  known (for instance
%from  CMB and  SNeIa analysis)  observations  of such  weak lensing  effects
%should  provide  valuable  information  on  the properties  of  the  density
%field.  Indeed, our  approach  explicitly  shows how  the  PDF $P(\Map)$  is
%related   to   the   $p-$point   correlation  functions   of   the   density
%field.  Although this  relation may  not  be directly  inverted an  analytic
%method like ours would allow a  fast investigation of the possible models of
%the density field which would already provide some interesting constraints.

\acknowledgements

We thank  Y. Mellier and L.  Van Waerbeke for fruitful  discussions.  We are
also very grateful to  B. Jain, U. Seljak and S. White  for the use of their
ray-tracing simulations  and to  K. Reblinsky for  providing us some  of the
data published in Reblinsky et al. (1999).

\appendix

\section{Order of the singularity of $\varphip(y)$}

As noticed in Valageas (2000a) the exponent $\omega_{s,c}$ of the singularity
of  $\varphic(y)$ (as defined  in (\ref{ys}))  translates into  the exponent
$\omega_{s,p}=  \omega_{s,c}-1/2$  for  the  projected  generating  function
$\varphip(y)$. However,  in the  cases encountered in  this article  we have
$\omega_{s,c}=-3/2$  for   both  the  quasi-linear   and  highly  non-linear
regimes. Then $\omega_{s,p}=  -2$ is an integer but  the generating function
$\varphip(y)$ is still singular  at the points $y_{s\pm}^{\rm proj}$ through
logarithmic  factors.  To see  this,  it is  convenient  to  take the  third
derivative  of  the  relation  (\ref{phiproj})  which  is  governed  by  the
singularity  at  $y_{s\pm}^{\rm  proj}$  and  diverges  for  $y  \rightarrow
y_{s\pm}^{\rm proj}$  (while the  lower derivatives of  $\varphip(y)$ remain
finite at $y_{s\pm}^{\rm proj}$). This yields: 
\be 
\varphip^{(3)}(y)=\int \d
\drad       \,      F(\drad)^3       \,       \psi_{\theta}(\drad)^2      \,
\varphic^{(3)}[y\,F(\drad)\psi_{\theta}(\drad)] .
\label{phiproj3}
\ee 
For $y \rightarrow y_{s\pm}^{\rm proj}$ the integral is dominated by the
values    of    $\drad$    around     the    point    where    the    factor
$F(\drad)\psi_{\theta}(\drad)$  is maximum, since  $\varphic^{(3)}$ diverges
as  $|y-y_{s\pm}^{\rm proj}|^{-3/2}$  at this  point. Thus,  we  obtain from
(\ref{phiproj3}):   
\be    
y   \rightarrow   y_{s\pm}^{\rm    proj}   :   \;
\varphip^{(3)}(y)   \sim   \int_{-\infty}^{\infty}   \d  \drad   \,   \left|
(1-\drad^2) y -  y_{s\pm}^{\rm proj} \right|^{-3/2} .  
\ee  
After the change
of  variable  $t=|y/(y-y_{s\pm}^{\rm  proj})|  \drad^2$  we  obtain:  
\be  
y \rightarrow y_{s\pm}^{\rm proj}  : \; \varphip^{(3)}(y) \sim \int_0^{\infty}
\frac{\d t}{\sqrt{t}}  \, (1+t)^{-3/2} \, \frac{|y|^{-1/2}}{|y-y_{s\pm}^{\rm
proj}|}  
\ee  
which  gives:  
\be  
y \rightarrow  y_{s\pm}^{\rm  proj}  :  \;
\varphip^{(3)}(y) \sim \frac{1}{|y-y_{s\pm}^{\rm proj}|} .  
\ee 
Finally, the integration of this relation leads to: 
\be 
y \rightarrow y_{s\pm}^{\rm proj}
:  \; \varphip(y)  \sim  (y-y_{s\pm}^{\rm proj})^2  \, \ln  |y-y_{s\pm}^{\rm
proj}| 
\ee 
where we only wrote the most singular term.

\end{document}